\documentclass[journal]{IEEEtran}

\usepackage{hyperref,doi,url}
\hypersetup{colorlinks=true, linkcolor=blue, breaklinks=true, urlcolor=blue}

\usepackage{amsmath,amssymb,amsfonts}
\usepackage{amsmath}
\usepackage{amssymb}

\usepackage[table]{xcolor}
\usepackage{graphicx}
\usepackage{amsmath}
\usepackage{amssymb}
\usepackage{amsthm}
\usepackage{bm}
\usepackage[compress]{cite}
\usepackage{enumitem}
\usepackage{dsfont}
\usepackage{bbm}
\usepackage{bbold}
\usepackage{indentfirst}
\usepackage{hhline}
\usepackage{mathtools}
\usepackage{commath}
\usepackage{yhmath}
\usepackage{extarrows}
\usepackage{subcaption}

\usepackage{tikz}
\usepackage{pgfplots}
\usepackage[prependcaption,colorinlistoftodos]{todonotes}

\usepackage{lineno}
\usepackage{multirow}
\usepackage{amsfonts}
\usepackage{textcomp}
\usepackage{stfloats}

\usepackage{tikz}
\usetikzlibrary{shapes, arrows.meta}
\tikzset{%
    >=latex,
	pursuer/.style = {fill=red!40, circle, radius=0.2cm, minimum size=0.2cm, inner sep=0},
	evader/.style = {fill=blue!40, regular polygon, regular polygon sides=3, minimum size=0.27cm, inner sep=0},
	point/.style = {fill=black, circle, radius=0.05cm, minimum size=0.05cm, inner sep=0},
}

\usepackage{centernot}
\usepackage{version,xspace}
\usepackage{environ}
\usepackage{blkarray}
\usetikzlibrary{automata,positioning,arrows,through}

\usepackage{threeparttable}

\usepackage{float}
\usepackage{indentfirst}
\usepackage{mathrsfs}
\usepackage[linesnumbered,ruled,vlined]{algorithm2e}

\usepackage{algorithmic}
\usepackage{tabularx}
\usepackage{soul}

\newtheorem{defi}{\textbf{Definition}}
\newtheorem{thom}{\textbf{Theorem}}
\newtheorem{asp}{\textbf{Assumption}}
\newtheorem{rek}{\textbf{Remark}}

\newtheorem{lema}{\textbf{Lemma}}

\newtheorem{pbm}{\textbf{Problem}}

\newcommand{\thomref}[1]{Theorem~\ref{#1}}

\newcommand{\pbmref}[1]{Problem \ref{#1}}

\newcommand{\lemaref}[1]{Lemma \ref{#1}}

\newcommand{\dnum}{N_p}
\newcommand{\anum}{N_e}
\newcommand{\dteam}{\mathscr{P}}
\newcommand{\ateam}{\mathscr{E}}
\newcommand{\miniturnradius}{\kappa}
\newcommand{\goal}{\Omega_{\rm goal}}
\newcommand{\play}{\Omega_{\rm play}}
\newcommand{\targetline}{\mathcal{T}}

\newcommand{\AR}{\mathbb{E}} 
\newcommand{\BAR}{\partial\AR} 
\newcommand{\CAR}{\overline{\AR}}

\newcommand{\interior}{\mathrm{int}}
\newcommand{\edgeset}{\mathcal{E}}
\newcommand{\graph}{\mathcal{G}}

\begin{document}

\title{Multiplayer Homicidal Chauffeur Reach-Avoid Games via Guaranteed Winning Strategies}

\author{Rui~Yan, Ruiliang Deng, Haowen Lai, Weixian Zhang, Zongying Shi, \IEEEmembership{Member,~IEEE,} and Yisheng Zhong
	\thanks{This work was supported by the National Natural Science Foundation of China under Grant 61374034.}
	\thanks{R. Yan, R. Deng, H. Lai, W. Zhang, Z. Shi, and Y. Zhong are with the Department
		of Automation, Tsinghua University, Beijing 100084, China. {\tt\small
			\{yr15,drl20,lhw19,wx-zhang17\} @mails.tsinghua.edu.cn} and {\tt\small\{szy,zys-dau\}
			@mail.tsinghua.edu.cn}}
}

\maketitle

\IEEEpeerreviewmaketitle
\begin{abstract}
  This paper studies a planar multiplayer Homicidal Chauffeur reach-avoid differential game, where each pursuer is a Dubins car and each evader has simple motion. The pursuers aim to protect a goal region cooperatively from the evaders. Due to the high-dimensional strategy space among pursuers, we decompose the whole game into multiple one-pursuer-one-evader subgames, each of which is solved in an analytical approach instead of solving Hamilton-Jacobi-Isaacs equations. For each subgame, an \emph{evasion region (ER)} is introduced, based on which a pursuit strategy guaranteeing the winning of a simple-motion pursuer under specific conditions is proposed. Motivated by the simple-motion pursuer, a strategy for a Dubins-car pursuer is proposed when the pursuer-evader configuration satisfies \emph{separation condition (SC)} and \emph{interception orientation (IO)}. The necessary and sufficient condition on capture radius, minimum turning radius and speed ratio to guarantee the pursuit winning is derived. When the IO is not satisfied (Non-IO), a heading adjustment pursuit strategy is proposed, and the condition to achieve IO within a finite time, is given. Then, a two-step pursuit strategy is proposed for the SC and Non-IO case. A non-convex optimization problem is introduced to give a condition guaranteeing the winning of the pursuer. A polynomial equation gives a lower bound of the non-convex problem, providing a sufficient and efficient pursuit winning condition. Finally, these pairwise outcomes are collected for the pursuer-evader matching. Simulations are provided to illustrate the theoretical results.
\end{abstract}



\section{Introduction}

\emph{Problem description and motivation:} Differential game theory provides a proper framework for analyzing the strategic interactions among multiple dynamical agents. Reach-avoid differential games, which consider two parties with conflicting objectives, have received significant attention in the past few years. However, due to the hardness of solving the Hamilton-Jacobi-Isaacs (HJI) equation, most of works focus on either complex dynamical models with numerical methods, or simple dynamical models with analytical methods. Motivated by the classical Homicidal Chauffeur game which involves nonlinear dynamics and has limited analytical results, this paper studies a planar multiplayer Homicidal Chauffeur reach-avoid differential game via guaranteed winning strategies in a pairwise and analytical way. In this game, a group of pursuers is used to protect a region cooperatively from a group of evaders, in which each pursuer is a Dubins car and each evader has simple motion. 

\emph{Literature review:} The first instances of reach-avoid differential games were developed in \cite{IMM-AMB-CJT:05,KM-JL:11,ZZ-RT-HH-CJT:12}, where one player aims to reach a predefined goal region, while avoiding adversarial circumstance induced by an opposing player. Building on these pioneers, a variety of variations have been proposed, such as multiple players \cite{MC-ZZ-CJT:17,RY-ZS-YZ:20-1,EG-DWC-AVM-MP:20}, time-varying targets and constraints \cite{JFF-MC-CJT-SSS:15}, analytical approach \cite{RY-ZS-YZ:19}, non-convex target sets \cite{VMA-EG-DC-MS-SCS:20} and perimeter defense \cite{DS-VK:20}. Such games encompass a huge number of adversarial scenarios in robotics and control \cite{JY-XD-QL-ZR:18,JY-XD-QL-JL-ZR:19,HL-YW-FLW-KPV:20}, such as safe motion-planning, collision avoidance, oil pipelines protection, and border protection. 

The current research into reach-avoid differential games has largely focused on the computation of the barrier, or called the boundary of the reach-avoid set, and optimal strategies. For complex dynamical models such as Dubins car \cite{MC-SLH-MSV-SB-CJT:18}, differential drive robot \cite{UR-RM:16} and double integrator dynamics, the analytical expressions of the barrier and optimal strategies are intractable due to the hardness of the associated HJI equation, for which a number of numerical tools based on grids have been proposed \cite{MC-SLH-MSV-SB-CJT:18,JFF-MC-CJT-SSS:15,MC-SB-JFF-CJT:19}. Unfortunately, these approaches suffer from the tradeoffs between execution time and accuracy, because the computational burden explodes as the grid scale increases. For simple dynamical models such as simple-motion model \cite{RI:65}, analytical barriers and optimal strategies have been computed for many different variants, including bounded environments \cite{RY-ZS-YZ:19,RY-ZS-YZ:20-1,EG-DWC-MP:20-2}, three- or high-dimensional game spaces \cite{RY-XD-ZS-YZ-FB:19,EG-DWC-MP:20,RY-ZS-YZ:20-2}, and time constraints \cite{RY-ZS-YZ:19-2}. However, the simple-motion model is a little restrictive, because the player is allowed to change its direction instantaneously, which is often infeasible in many control applications such as safe planning for autonomous vehicles and border guarding by robots with minimum turning radius. As discussed in \cite{RY-ZS-YZ:20-1,MP-SC:19}, the analytical methods for reach-avoid differential games with complex dynamical models are urgently needed.  

This paper considers a variant of the classical Homicidal Chauffeur game, called \emph{Homicidal Chauffeur reach-avoid differential games}, in which the chauffeurs or pursuers (Dubins-car models)  aim to protect a region by capturing the pedestrians or evaders (simple-motion models \cite{RI:65}). The Homicidal Chauffeur game was initially invented by Isaacs in \cite{RI:65} and then a systematic description of the solution structure was presented in \cite{AWM:71} by Breakwell and Merz. The steps to obtain the full solution for the parameter range in the heart of the speed ratio and capture radius parameter space, are quite complex and also fantastic. We refer the interested readers to the references \cite{MP-SC:19} and \cite{VSP-VLT:11} for a thorough understanding of history and modern studies on Homicidal Chauffeur game. Compared with the classical Homicidal Chauffeur game where the capture is the unique goal, Homicidal Chauffeur reach-avoid games are more complicated and have more practical significance, because the pedestrians not only aim to avoid the capture but also strive to reach a goal region. Two terminal conditions, the capture of evader or arrival in the goal region, make it difficult to extend the analytical method in  \cite{RI:65} and \cite{AWM:71} for the variant we are considering in this paper, as this method requires backward integration from terminal manifold and different backward trajectories may produce complicated singular surfaces, for which there exist no systematic analysis methods \cite{JL:12}.

There are several attempts to analytically study adversarial games with Dubins cars. For example, Bopardikar \emph{et al.} \cite{SDB-FB-JPH:09} proposed a multi-phase strategy to confine an evader into a bounded region formed by the pursuers for a cooperative Homicidal Chauffeur game. In \cite{DWO-ARG:16}, the authors studied the dominance regions for the Homicidal Chauffeur game. The work \cite{IE-PT-MP:15} introduced a reversed Homicidal Chauffeur game, called \emph{Suicidal Pedestrian differential game}, and derived the winning regions and optimal strategies. In \cite{SC-MP-RM:17}, the optimal control of a Dubins car was discussed for the Homicidal Chauffeur game with a stationary evader. However, to our best knowledge, there is no existing literature to analytically study the reach-avoid differential games with Dubins cars, especially Homicidal Chauffeur reach-avoid differential games.

\emph{Contributions:} In this paper, we study the analytical cooperative strategies for multiple Dubins-car pursuers to protect a region in the plane from multiple simple-motion evaders. Compared with \cite{MC-ZZ-CJT:17,MC-SB-JFF-CJT:19,MC-SLH-MSV-SB-CJT:18,JFF-MC-CJT-SSS:15,EG-DWC-MP:20-2,EG-DWC-AVM-MP:20,EG-DWC-MP:20,VMA-EG-DC-MS-SCS:20,RY-ZS-YZ:19-2,RY-XD-ZS-YZ-FB:19,RY-ZS-YZ:20-2,RY-ZS-YZ:20-1,RY-ZS-YZ:19}, we combine the advantages of analytical methods in accuracy and computational efficiency, and Dubins-car models in broader applications. The main contributions are as follows:
\begin{enumerate}[label=(\roman*)]
    \item For each subgame with one pursuer and one evader, when the minimum turning radius is zero, i.e., a simple-motion pursuer is considered, a pursuit strategy based on the evasion region (ER), is proposed such that the pursuer can protect the goal region from the evader under specific initial configurations.
    \item Building on the above, when the minimum turning radius is positive, i.e., a Dubins-car pursuer is considered, a pursuit strategy is proposed when the pursuer-evader configuration satisfies \emph{separation condition (SC)} and \emph{interception orientation (IO)}. Then, the necessary and sufficient condition on capture radius, minimum turning radius and speed ratio to guarantee the pursuit winning, is derived.
    \item If the configuration does not satisfy IO (i.e., Non-IO), a heading adjustment pursuit strategy is proposed. Under this strategy, a sufficient condition on capture radius, minimum turning radius and speed ratio for steering the configuration into IO after a finite time, is presented. Furthermore, an upper bound for the heading adjustment time is given. 
    \item For the SC and Non-IO case, a two-step pursuit strategy is proposed by merging the heading adjustment strategy and the strategy in the case of SC and IO. A sufficient condition on capture radius, minimum turning radius and speed ratio, as well as the optimal value of a non-convex optimization problem, for the guaranteed pursuit winning, is given. A lower bound of the non-convex optimization problem is computed by solving a sextic equation, thus providing a sufficient and efficient pursuit winning condition. Finally, all pairwise outcomes are collected for the maximum pursuer-evader matching and thus a receding horizon pursuit strategy is proposed. 
\end{enumerate}

\emph{Paper organization:} We introduce the Homicidal Chauffeur reach-avoid differential games in Section \ref{sec:problem-statement}, including problem description, information structure and assumptions. In Section \ref{sec:simple-motion-pursuer}, the case where one simple-motion pursuer plays against one evader, is discussed. Based on it, Section \ref{sec:Dubins-car-pursuer} presents the main results of one Dubins-car pursuer against one simple-motion evader, including pursuit winning strategy and heading adjustment strategy. In Section \ref{sec:multiplayer}, a receding horizon pursuit strategy based on pairwise outcomes and maximum matching is proposed. Numerical results are presented in Section \ref{sec:simulation}, and we conclude the paper in Section \ref{sec:conclusion}.

\emph{Notation:} Let $\mathbb{R}$, $\mathbb{R}_{>0}$ and $\mathbb{R}_{\ge0}$ be the set of reals, positive reals and nonnegative reals, respectively. Let $\mathbb{R}^n$ be the set of $n$-dimensional real column vectors and $\enVert[0]{\cdot}_2$ be the Euclidean norm. All vectors in this paper are column vectors. Let $\bm{0}$ denote the zero vector whose dimension will be  clear from the context. Denote the unit desk in $\mathbb{R}^n$ by $\mathbb{S}^{n-1}$, i.e., $\mathbb{S}^{n-1}=\{ \bm{u}\in\mathbb{R}^n\, |\, \|\bm{u}\|_2 \leq 1 \}$. For any set $S\subset\mathbb{R}^n$, let $\interior(S)$, $\partial S$ and $\overline{S}$ be the \emph{interior}, \emph{boundary} and \emph{closure} of $S$, respectively. Let $\textup{sgn}(\cdot)$ be the sign function. 


\section{Problem Statement}\label{sec:problem-statement}
\subsection{Homicidal Chauffeur Reach-Avoid Differential Games}
\begin{figure}[tbp]
    \centering
    \begin{tikzpicture}
        \node[pursuer, label=right:$\bm{x}_{P_1}$] at (0, 0) {};
        \draw[thick, densely dotted, red!40] (0, 0) arc (0:-50:0.5cm);
        \draw[thick, red!40, ->] (0, 0) arc (0:50:0.5cm);
        
        \node[pursuer, label=right:$\bm{x}_{P_2}$] at (1, 1) {};
        \draw[thick, densely dotted, red!40] (1, 1) arc (-180:-120:0.7cm);
        \draw[thick, red!40, ->] (1, 1) arc (180:130:0.7cm);

        \node[pursuer, label=left:$\bm{x}_{P_3}$] at (-2, 0.5) {};
        \draw[thick, densely dotted, red!40] (-2, 0.5) arc (-180:-120:0.5cm);
        \draw[thick, red!40, ->] (-2, 0.5) arc (180:130:0.5cm);
        
        \node[evader, label=left:$\bm{x}_{E_1}$] at (-0.5, 1.8) {};
        \draw[thick, densely dotted, blue!40] (-0.3, 2.3) -- (-0.5, 1.8);
        \draw[thick, blue!40, ->] (-0.5, 1.8) -- (-0.7, 1.3);

        \node[evader, label=right:$\bm{x}_{E_2}$] at (1.7, 2) {};
        \draw[thick, densely dotted, blue!40] (1.6, 2.5) -- (1.7, 2);
        \draw[thick, blue!40, ->] (1.7, 2) -- (1.8, 1.5);
        
        \draw [fill, green!30] (-3.4, -0.5) rectangle (2.9, -1.5);
        \draw [black!50!green, very thick] (-3.5, -0.5) -- (3, -0.5);

        \node [anchor=west] at (3, -0.5) {$\targetline$};
        \node at (0, -1) {$\goal$};
        \node [anchor=west] at (3, 1.25) {$\play$};

    \end{tikzpicture}
    \caption{Multiplayer Homicidal Chauffeur reach-avoid differential games with Dubins-car pursuers (red circles) and simple-motion evaders (blue triangles).}
    \label{fig:game-description}
\end{figure}
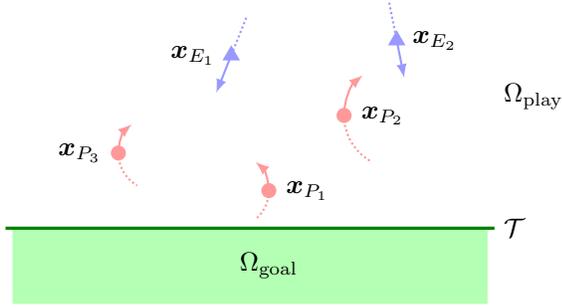

Consider a reach-avoid differential game in the plane $\mathbb{R}^2$ with $\dnum+\anum$ players, where there are $\dnum$ pursuers  $\dteam=\{P_1,\dots,P_{\dnum}\}$ and $\anum$ evaders $\ateam=\{E_1,\dots,E_{\anum}\}$. The players are assumed to be mass points and their dynamics are described by Homicidal Chauffeur models \cite{RI:65}. Each pursuer $P_i\in\dteam$ is modeled as a Dubins car: 
\begin{equation}\label{eq:pursuer_car}
    \begin{aligned}
        \dot{x}_{P_i}&=v_{P_i}\cos\theta_{P_i},& x_{P_i}(0)&=x_{P_i0},\\
        \dot{y}_{P_i}&=v_{P_i}\sin\theta_{P_i},&y_{P_i}(0)&=y_{P_i0},\\
        \dot{\theta}_{P_i}&=v_{P_i}u_{P_i}/\miniturnradius_i,&\theta_{P_i}(0)&=\theta_{P_i0},
    \end{aligned}
\end{equation}
where $\bm{x}_{P_i}=[x_{P_i}, y_{P_i}]^\top\in\mathbb{R}^2$, $\theta_{P_i}\in[0,2\pi)$ and $u_{P_{i}}\in\mathbb{S}^0$ are pursuer $P_i$'s position, heading and control input respectively, and $v_{P_i}\in\mathbb{R}_{>0}$ and $\miniturnradius_i\in\mathbb{R}_{>0}$ are the constant speed and minimum turning radius of $P_i$ respectively. The initial position and heading of $P_i$ are $\bm{x}_{P_i0}=[x_{P_i0}, y_{P_i0}]^\top\in\mathbb{R}^2$ and $\theta_{P_i0}\in[0,2\pi)$, respectively. Each evader $E_j\in\ateam$ has simple motion:
\begin{equation}\label{eq:evader_simple}
    \begin{aligned}
        \dot{x}_{E_j}&=v_{E_j}u_{E_j}^x,&x_{E_j}(0)&=x_{E_j0},\\ 
        \dot{y}_{E_j}&=v_{E_j}u_{E_j}^y,&y_{E_j}(0)&=y_{E_j0},
    \end{aligned}
\end{equation}
where $\bm{x}_{E_j}=[x_{E_j},y_{E_j}]^\top\in\mathbb{R}^2$ and  $\bm{u}_{E_j}=[u_{E_j}^x,u_{E_j}^y]^\top\in\mathbb{S}^1$ are evader $E_j$'s position and control input respectively, and $v_{E_j}\in\mathbb{R}_{>0}$ is the constant speed of $E_j$. The initial position of $E_j$ is $\bm{x}_{E_j0}=[x_{E_j0},y_{E_j0}]^\top\in\mathbb{R}^2$. The goal region $\goal\subset\mathbb{R}^2$ is assumed to be a half-plane, and its boundary is denoted by $\targetline\subset\mathbb{R}^2$, as shown in Fig. \ref{fig:game-description}. The complementary set of $\goal$ in $\mathbb{R}^2$ is called play region $\play$. We take
\begin{equation*}
    \begin{aligned}
        \goal&=\{\bm{x}\in\mathbb{R}^2\,|\,g(\bm{x})\le0\},\\
        \targetline&=\{\bm{x}\in\mathbb{R}^2\,|\,g(\bm{x})=0\},\\
        \play&=\{\bm{x}\in\mathbb{R}^2\,|\,g(\bm{x})>0\},
    \end{aligned}
\end{equation*}
where $g(\bm{x}):=[0,1]\bm{x}$.


Suppose that the pursuer $P_i$ has capture radius $r_i>0$. The evader is captured as soon as its distance to at least one pursuer becomes equal to or less than the corresponding capture radius. Assume that the number of pursuers remains constant, and the pursuers chase the evaders until no evaders in $\play$, that is, the game ends in this situation. 


The evaders $\ateam$ (evasion team) strive to send as many evaders as possible into the goal region $\goal$ before being captured, while the pursuers $\dteam$ (pursuit team) try to protect $\goal$ by capturing as many evaders as possible in the play region $\play$. This paper aims at designing a receding horizon pursuit strategy for the pursuit team which can give a lower bound of the number of guaranteed captured evaders.

\subsection{Information Structure and Assumptions}
In the differential games, the information available to each player plays an important role in determining game outcomes. This paper considers a nonanticipative information structure (see for example \cite{RJE-NJK:72}, \cite{IMM-AMB-CJT:05}). Under this information structure, the pursuit team makes decisions about its current input with the information of all players' current states, plus the evasion team's current control input. While the evasion team is at a slight disadvantage under this information structure, at a minimum it has access to the information of all players' current states, because the pursuit team must declare its strategy before the evasion team chooses a specific input and thus the evasion team can determine the response of the pursuit team to any input signal. Thus, the game formulated here is an instantiation of the Stackelberg game \cite{TB-GJO:99}. 

Assume that the initial positions of all players satisfy the following conditions, which can focus our attention on the main situations and remove some technical problems (eg., two pursuers or evaders initially lie at the same position).
\begin{asp}[Initial deployment]\label{asp:initial-deployment}
The initial positions of all players satisfy the following four conditions:
\begin{enumerate}
    \item $\enVert[0]{\bm{x}_{P_i0}-\bm{x}_{P_j0}}_2>0$ for all $P_i,P_j\in\dteam,P_i\neq P_j$;
    \item $\enVert[0]{\bm{x}_{E_i0}-\bm{x}_{E_j0}}_2>0$ for all $E_i,E_j\in\ateam,E_i\neq E_j$;
    \item $\enVert[0]{\bm{x}_{E_j0}-\bm{x}_{P_i0}}_2>r_i$ for all $P_i\in\dteam,E_j\in\ateam$;
    \item $\bm{x}_{P_i0}\in\mathbb{R}^2$ for all $P_i\in\dteam$ and $\bm{x}_{E_j0}\in\play$ for all $E_j\in\ateam$.
\end{enumerate}
\end{asp}

 In the Homicidal Chauffeur game, the chauffeur is faster. Similarly, we focus on the faster pursuer case.

\begin{asp}[Speed ratio]
Suppose that the speed ratio $\alpha_{ij}=v_{P_i}/v_{E_j}>1$ for all $P_i\in\dteam$ and $E_j\in\ateam$.
\end{asp}



\section{One vs. One Games: Simple-Motion Pursuer}\label{sec:simple-motion-pursuer}

It is intractable to analyze the whole game directly due to the high-dimensional strategy space and complex cooperation. We here overcome this intractability by decomposing the game into multiple one-puruser-one-evader subgames. Then, we look into these subgames and collect the pairwise outcomes for the pursuer-evader matching.  Next, these subgames will be discussed.

 It is known that the strategies in the Homicidal Chauffeur games are quite complex \cite{AWM:71}. The strategies in the Homicidal Chauffeur reach-avoid games become more complicated, because the evader not only tries to escape but also strives to enter a goal region. We will start with the simple-motion pursuer, and propose a pursuit strategy under which a guaranteed pursuit winning condition is derived. Then, these insights will be used for the Dubins-car pursuer in the next section. Suppose that $\kappa_i$ is small enough. Each pursuer $P_i\in\dteam$ in \eqref{eq:pursuer_car} becomes simple motion: 
\begin{equation}\label{eq:pursuer_simple}
    \begin{aligned}
        \dot{x}_{P_i}&=v_{P_i}u_{P_i}^x,&x_{P_i}(0)&=x_{P_{i0}},\\ 
        \dot{y}_{P_i}&=v_{P_i}u_{P_i}^y,&y_{P_i}(0)&=y_{P_{i0}},
    \end{aligned}
\end{equation}
where $\bm{u}_{P_i}=[u_{P_i}^x,u_{P_i}^y]^\top\in\mathbb{S}^1$ is the  control input of $P_i$. In this section, we focus on \eqref{eq:pursuer_simple} for the pursuer.  

\subsection{Evasion Region}

First, we revisit a class of potential functions introduced in \cite{RY-XD-ZS-YZ-FB:19} with a little revision, that is, we drop off the capture radius.

\begin{defi}[Potential function]
Given any $\bm{x}_{P_i}$ and $\bm{x}_{E_j}$ at time $t\ge0$, define the potential function $f_{ij}:\mathbb{R}^2\times\mathbb{R}_{\ge0}\to\mathbb{R}$ associated with $P_i$ and $E_j$ as follows 
\begin{equation}
    f_{ij}(\bm{x},t)=\enVert[0]{\bm{x}-\bm{x}_{P_i}}_2-\alpha_{ij}\enVert[0]{\bm{x}-\bm{x}_{E_j}}_2,
\end{equation}
whose gradient with respect to $\bm{x}$ is denoted by $\nabla_{\bm{x}} f_{ij}(\bm{x},t)\in\mathbb{R}^2$, and given by
\begin{equation}
    \nabla_{\bm{x}} f_{ij}(\bm{x},t)=\frac{\bm{x}-\bm{x}_{P_i}}{\enVert[0]{\bm{x}-\bm{x}_{P_i}}_2}-\alpha_{ij}\frac{\bm{x}-\bm{x}_{E_j}}{\enVert[0]{\bm{x}-\bm{x}_{E_j}}_2},
\end{equation}
when $\bm{x}\neq \bm{x}_{P_i}$ and $\bm{x}\neq \bm{x}_{E_j}$.
\end{defi}

We consider the concept introduced in Section 6.7 in Isaacs' book \cite{RI:65} as follows.

\begin{defi}[Evasion region]\label{defi:AR}
Given $\bm{x}_{P_i}$ and $\bm{x}_{E_j}$, the evasion region (ER) $\AR_{ij}$, is the set of positions in $\mathbb{R}^2$ that $E_j$ can \emph{reach} before $P_i$ for any control input $\bm{u}_{P_i}\in\mathbb{S}^1$, i.e., $\AR_{ij}=\{\bm{x}\in\mathbb{R}^2\,|\,f_{ij}(\bm{x},t)>0\}$.
\end{defi}

Unless for clarity, we will suppress the dependence on $t$. Let $\BAR_{ij}$ and $\CAR_{ij}$ denote the boundary and closure of $\AR_{ij}$, respectively. Thus, they can be respectively formulated as
\begin{equation*}\begin{aligned}\label{eq:closureofAR-1v1}
    \BAR_{ij}&=\big\{\bm{x}\in\mathbb{R}^2\,|\,f_{ij}(\bm{x},t)=0\big\},\\
    \CAR_{ij}&=\big\{\bm{x}\in\mathbb{R}^2\,|\,f_{ij}(\bm{x},t)\ge0\big\}.
\end{aligned}\end{equation*}
By Definition \ref{defi:AR}, the ER $\AR_{ij}$ is the interior of a circle with
\begin{equation*}
\textup{center: } \frac{\alpha_{ij}^2\bm{x}_{E_j}-\bm{x}_{P_i}}{\alpha^2_{ij}-1}, \quad\ \textup{radius: }\frac{\alpha_{ij}\enVert[0]{\bm{x}_{P_i}-\bm{x}_{E_j}}_2}{\alpha_{ij}^2-1}.
\end{equation*}

\subsection{Pursuit Winning Strategy}

The distance between two points $\bm{x}\in\mathbb{R}^2$ and $\bm{y}\in\mathbb{R}^2$ is defined as $\rho(\bm{x}, \bm{y})=\enVert[0]{\bm{x}-\bm{y}}_2$, the distance between a point $\bm{x}\in\mathbb{R}^2$ and a non-empty set $\mathcal{M}\subset\mathbb{R}^2$ is defined as $\rho(\bm{x}, \mathcal{M})=\inf_{\bm{y}\in\mathcal{M}}\enVert[0]{\bm{x}-\bm{y}}_2$, and the distance between two non-empty sets $\mathcal{M}_1$ and $\mathcal{M}_2$ is
\begin{equation*}
\begin{aligned}
&\rho(\mathcal{M}_1,\mathcal{M}_2)\\
&=
\begin{cases}
\underset{\bm{x}\in\mathcal{M}_1,\bm{y}\in\mathcal{M}_2}{\inf}\enVert[0]{\bm{x}-\bm{y}}_2, \ \ \,  \textup{if }\interior(\mathcal{M}_1) \cap \interior(\mathcal{M}_2)=\emptyset,\\
-\infty,\qquad \qquad\qquad \quad\ \ \ \  \textup{otherwise}.
\end{cases}
\end{aligned}
\end{equation*}
Let $\rho_{\targetline}:\mathbb{R}^2\to\mathbb{R}$ be the signed distance function to $\targetline$:
\begin{equation*}
\rho_{\targetline}(\bm{x}):=
\begin{cases}
\rho(\bm{x},\targetline), & \bm{x}\in\play,\\
-\rho(\bm{x},\targetline), & \bm{x}\in\goal.
\end{cases}
\end{equation*}
For any $\bm{x}_{P_i}$ and $\bm{x}_{E_j}$, according to the radius of $\AR_{ij}$, let
\begin{equation*}
   \Upsilon(\bm{x}_{P_i},\bm{x}_{E_j})=\Big[
    0, - \frac{\alpha_{ij} \enVert[0]{ \bm{x}_{P_i} - \bm{x}_{E_j} }_2}{\alpha_{ij}^2-1}
\Big]^\top.
\end{equation*}


Next, a critical point in $\CAR_{ij}$ is introduced.

\begin{defi}[Interception point]\label{defi:interception-point}
Given any $\bm{x}_{P_i}$ and $\bm{x}_{E_j}$, let the interception point $I_{ij}$ given by $\bm{x}_{I_{ij}}=[x_{I_{ij}},y_{I_{ij}}]^\top$ be the unique point in $\CAR_{ij}$ that has the minimum signed distance to $\targetline$, i.e., $\bm{x}_{I_{ij}}=\arg\min_{\bm{x}\in\CAR_{ij}}\rho_{\targetline}(\bm{x})$ and is given by
\begin{equation}\label{eq:XI}
    \bm{x}_{I_{ij}}=\frac{\alpha_{ij}^2\bm{x}_{E_j}-\bm{x}_{P_i}}{\alpha^2_{ij}-1}+\Upsilon(\bm{x}_{P_i},\bm{x}_{E_j}).
\end{equation}
\end{defi}

For clarity, the indexes $i$ and $j$ are omitted hereinafter. Note that if $\bm{x}_P\neq\bm{x}_E$, then $\bm{x}_P\neq\bm{x}_I$ and $\bm{x}_E\neq\bm{x}_I$ hold. The following theorem presents a \emph{pursuit winning strategy}.

\begin{thom}[Pursuit winning strategy]\label{thom:dwinstra-simple}
Consider $P\in\dteam$ and $E\in\ateam$ under the model \eqref{eq:pursuer_simple} and \eqref{eq:evader_simple}, respectively. Suppose that $\rho(\CAR,\goal)\ge0$. If $P$ adopts the feedback strategy $\bm{u}_{P}=\frac{\bm{x}_{I}-\bm{x}_{P}}{\enVert[0]{\bm{x}_{I}-\bm{x}_{P}}_2}$, then $P$ guarantees that $\CAR$ does not approach $\goal$, that is, 
\begin{equation}\label{eq:no-close-inequality}
    \od{}{t} \rho\big( \CAR, \goal \big) \ge 0,
\end{equation}
for any $\bm{u}_{E}\in\mathbb{S}^1$. Moreover, the equality holds if and only if $E$ adopts the feedback strategy $\bm{u}_{E}=\frac{\bm{x}_{I}-\bm{x}_{E}}{\enVert[0]{\bm{x}_{I}-\bm{x}_{E}}_2}$.
\end{thom}
\begin{proof}
    We have $\rho(\CAR,\goal)\ge0$. Note that $\CAR$ is strictly convex and $\goal$ is convex. Thus, according to Definition~\ref{defi:interception-point}, at any time $t\ge 0$, for the interception point $\bm{x}_I(t)$, there exists a unique point $\bm{y}_I(t)$ in $\goal$, such that the pair $(\bm{x}_I(t),\bm{y}_I(t))$ is the unique solution of the convex problem
		\begin{equation*}
		\begin{aligned}
		& \underset{(\bm{x},\bm{y})\in\mathbb{R}^2\times\mathbb{R}^2}{\textup{minimize}}
			&& \rho(\bm{x},\bm{y})\\
			&\textup{ subject to}&& f(\bm{x},t)\ge 0,\quad g(\bm{y})\le 0.
		\end{aligned}
		\end{equation*}
		In the above problem, Since $\bm{x}_P\neq \bm{x}_E$, then $f(\cdot,\cdot)$ and $g(\cdot)$ are smooth functions related to the closure of the time-varying ER  $\CAR$, and the constant goal region $\goal$, respectively, i.e.,
		\begin{equation*}
		\begin{aligned}
			f(\bm{x},t)\ge 0&\iff \bm{x}\in\CAR \text{ at time $t$},\\
			g(\bm{y})\le 0&\iff \bm{y}\in\goal.
		\end{aligned}
		\end{equation*}
		
		According to the \emph{Karush-Kuhn-Tucker (KKT) conditions}, at any time $t\ge 0$, the solution $\left(\bm{x}_I(t), \bm{y}_I(t) \right)$ satisfies
		\begin{equation}\label{eq:kkt}
        \begin{aligned}
        	&\bm{0}=\nabla_{\bm{x}} \rho	\left(\bm{x}_I(t),\bm{y}_I(t)\right)+ \lambda_1(t)\cdot \nabla_{\bm{x}} f\left(\bm{x}_I(t),t\right),\\
        	&\bm{0}=\nabla_{\bm{y}} \rho	\left(\bm{x}_I(t),\bm{y}_I(t)\right) + \lambda_2(t)\cdot\nabla_{\bm{y}} g\left(\bm{y}_I(t)\right),\\
        	& f\left(\bm{x}_I(t),t\right)\ge 0, \quad \qquad \ g\left(\bm{y}_I(t)\right)\le 0,\\
        	&\lambda_1(t)\le 0, \quad \ \ \, \qquad \qquad \lambda_2(t)\ge 0,\\
        	& \lambda_1(t)f\left(\bm{x}_I(t),t\right)=0,\quad  \lambda_2(t)g\left(\bm{y}_I(t)\right)=0,
        \end{aligned}
		\end{equation}
		where $\lambda_1(t)\in\mathbb{R}$ and $\lambda_2(t)\in\mathbb{R}$ are the Lagrange multipliers, and $\nabla_{\bm{x}}$ and $\nabla_{\bm{y}}$ represent the gradient operators with respect to $\bm{x}$ and $\bm{y}$ respectively. Thus, the time derivative of the distance between two sets $\CAR$ and $\goal$ can be computed by
		\begin{equation*}
		\begin{aligned}
			 &\dod{}{t}\rho\left( \CAR, \goal \right)\\
			&=  \dod{}{t}\rho\left( \bm{x}_I(t),\bm{y}_I(t) \right)\\
			 &=  \nabla^\top_{\bm{x}}\rho\dod{\bm{x}_I(t)}{t} + \nabla^\top_{\bm{y}}\rho\dod{\bm{y}_I(t)}{t}\\
			&\xlongequal{\eqref{eq:kkt}}  -\lambda_1(t) \nabla^\top_{\bm{x}}f \dod{\bm{x}_I(t)}{t} - \lambda_2(t) \nabla^\top_{\bm{y}}g \dod{\bm{y}_I(t)}{t}.
		\end{aligned}
		\end{equation*}
		
		Apparently, $\bm{y}_I(t)$ is always at the boundary of $\goal$, i.e., $g( \bm{y}_I(t) )\equiv 0$. Thus, $\nabla^\top_{\bm{y}}g\od{\bm{y}_I(t)}{t} = 0$ holds. On the other hand, the proof of Theorem 3.1 in \cite{RY-XD-ZS-YZ-FB:19} shows that if $P$ adopts the feedback strategy $\bm{u}_{P} = \frac{\bm{x}_{I} - \bm{x}_{P}}{\enVert[0]{\bm{x}_{I} - \bm{x}_{P}}_2}$, then $\nabla^\top_{\bm{x}}f\od{\bm{x}_I(t)}{t}\ge 0$ holds. Thus $\od{}{t}\rho\left(\CAR, \goal \right)\ge 0$ at any time $t$. The conclusion follows from Theorem 3.1 in \cite{RY-XD-ZS-YZ-FB:19}.
\end{proof}

From now on, we refer to a pursuit strategy as a pursuit winning strategy if it can guarantee both $\rho(\CAR,\goal)\ge0$ and \eqref{eq:no-close-inequality}, because in this situation the evader can never reach $\goal$. This situation is also referred to as a pursuit winning.

\section{One vs. One Games: Dubins-Car Pursuer}\label{sec:Dubins-car-pursuer}
With the insights about simple-motion pursuer, we are now ready to analyze the subgame between one Dubins-car pursuer and one simple-motion evader.

\subsection{Pursuit Winning Strategy for SC and IO}\label{subsec:PWS-IO}



Before presenting the main results, we first define several notations. Let $X(t)=(\bm{x}_{P}(t),\theta_{P}(t),\bm{x}_{E}(t))\in\mathbb{R}^2\times[0,2\pi)\times\mathbb{R}^2$ denote the state of the game at time $t$. Define two mappings $F:\mathbb{R}^2\times[0,2\pi)\times\mathbb{R}^2\to\mathbb{R}^2$ and $G:\mathbb{R}^2\times[0,2\pi)\times\mathbb{R}^2\to\mathbb{R}$, where $ F(X)=[F_x(X),F_y(X)]^\top$, and $F_x(X)$, $F_y(X)$ and $G(X)$ are given by \eqref{eq:defensewinparameter1}. 
Let $h(\alpha)$ be the optimal value of the following problem:
\begin{equation}\label{eq:alphasolutionzero}
    \begin{aligned}
       &\underset{x\in\mathbb{R},\,y\in\mathbb{R}}{\textup{maximize}}&& \frac{y+\alpha}{2\alpha y+\alpha^2+1}+\frac{\alpha x(y+\alpha)}{(2\alpha y+\alpha^2+1)^{\frac{3}{2}}} \\
       & \textup{subject to}&& x^2+y^2=1.
    \end{aligned}
\end{equation}

Next, we introduce two conditions to classify the states into several classes, and analyze them separately.

\begin{figure*}[ht]
\begin{equation}\label{eq:defensewinparameter1}
\begin{aligned}
    &F_x(X)=\frac{\kappa(y_{P}-y_{E})\big(\alpha\enVert[0]{\bm{x}_{P}-\bm{x}_{E}}_2+(y_{P}-y_{E})\big)}{\enVert[0]{\bm{x}_{P}-\bm{x}_{E}}_2^2\big((\alpha^2+1)\enVert[0]{\bm{x}_{P}-\bm{x}_{E}}_2+2\alpha(y_{P}-y_{E})\big)},\\
    &F_y(X)=\frac{-\kappa(x_{P}-x_{E})\big(\alpha\enVert[0]{\bm{x}_{P}-\bm{x}_{E}}_2+(y_{P}-y_{E})\big)}{\enVert[0]{\bm{x}_{P}-\bm{x}_{E}}_2^2\big((\alpha^2+1)\enVert[0]{\bm{x}_{P}-\bm{x}_{E}}_2+2\alpha(y_{P}-y_{E})\big)},\\
    &G(X)=\frac{-\kappa\alpha(x_{P}-x_{E})\big(\alpha\enVert[0]{\bm{x}_{P}-\bm{x}_{E}}_2+(y_{P}-y_{E})\big)}{\enVert[0]{\bm{x}_{P}-\bm{x}_{E}}_2^{\frac{3}{2}}\big((\alpha^2+1)\enVert[0]{\bm{x}_{P}-\bm{x}_{E}}_2+2\alpha(y_{P}-y_{E})\big)^{\frac{3}{2}}}.
\end{aligned}
\end{equation}
\end{figure*}

\begin{defi}[Separation condition]
A state $X$ satisfies separation condition (SC) if $\rho(\CAR,\goal)\ge0$ holds.
\end{defi}

\begin{defi}[Interception orientation]\label{def:interception-orientation}
For any given state $X=(\bm{x}_{P},\theta_{P},\bm{x}_{E})$, the interception angle $\theta_I\in[0,2\pi)$ is defined by 
\begin{equation*}\begin{aligned}
\cos\theta_{I}=\frac{x_{I}-x_{P}}{\enVert[0]{\bm{x}_{I}-\bm{x}_{P}}_2},\ \sin\theta_{I}=\frac{y_{I}-y_{P}}{\enVert[0]{\bm{x}_{I}-\bm{x}_{P}}_2}.
\end{aligned}\end{equation*}
Then, the state $X$ satisfies interception orientation (IO) if $P$'s heading  is equal to the interception angle, i.e., $\theta_{P}=\theta_{I}$.
\end{defi}

We now present the pursuit winning strategy when the state $X$ satisfies SC and IO.

\begin{thom}[Pursuit winning strategy for SC and IO]\label{thom:dwinstra-car-zero}
Consider $P\in\dteam$ and $E\in\ateam$ under the model \eqref{eq:pursuer_car} and \eqref{eq:evader_simple}, respectively. Let a state $X$ which satisfies SC and IO. If $P$ adopts the feedback strategy
\begin{equation}\label{eq:dwincontrolzero}
u_{P}=F(X)^\top \bm{u}_{E}+G(X),
\end{equation}
then the SC and IO hold all the time before the capture and $P$ also guarantees that $\CAR$ does not approach $\goal$, i.e., 
\begin{equation}\label{eq:odzeroradius}
    \dod{}{t} \rho( \CAR, \goal ) \ge 0,
\end{equation}
for any $\bm{u}_{E}\in\mathbb{S}^1$, if and only if the capture radius $r$, minimum turning radius $\miniturnradius$ and speed ratio $\alpha$ satisfy
\begin{equation}\label{eq:parameterrelazero}
    r-\kappa h(\alpha)\ge0.
\end{equation}
Moreover, the equality in \eqref{eq:odzeroradius} holds if and only if $E$ adopts the feedback strategy $\bm{u}_{E}=\frac{\bm{x}_{I}-\bm{x}_{E}}{\enVert[0]{\bm{x}_{I}-\bm{x}_{E}}_2}$.
\end{thom}
\begin{proof}
Note that the state $X$ satisfies SC. Thus, according to Theorem \ref{thom:dwinstra-simple}, by starting from $X$, if the positions of $P$ and $E$ evolves along the following dynamics
\begin{equation}\label{eq:onlyAdynamics}
    \begin{aligned}
        \Dot{\bm{x}}_E = v_E \bm{u}_E,\quad \Dot{\bm{x}}_P = v_P \frac{\bm{x}_I-\bm{x}_P}{\enVert[0]{\bm{x}_I-\bm{x}_P}_2},
    \end{aligned}
\end{equation}
then the SC and IO hold all the time and $P$ can guarantee that $\CAR$ does not approach $\goal$, i.e., \eqref{eq:odzeroradius} holds for any $\bm{u}_{E}\in\mathbb{S}^1$. Therefore, our goal is to design a pursuit strategy such that the dynamics \eqref{eq:onlyAdynamics} is satisfied unless $E$ is captured by $P$.

Note that \eqref{eq:onlyAdynamics} can be regarded as a nonlinear dynamical system with a unique control input $\bm{u}_E$. For any given $\bm{u}_E$, the signed curvature \cite{AG-EA-SS:17} of $P$'s trajectory under the dynamics \eqref{eq:onlyAdynamics} can be computed by 
\begin{equation}\label{eq:curvaturezero}
  \dod{\theta_P}{s_P}=\frac{\ddot{y}_P\dot{x}_P-\ddot{x}_P\dot{y}_P}{(\dot{x}_P^2+\dot{y}_P^2)^{\frac{3}{2}}}=\frac{\ddot{y}_P\dot{x}_P-\ddot{x}_P\dot{y}_P}{v_P^3},  
\end{equation}
where $s_P\in\mathbb{R}_{\ge0}$ is $P$'s moving distance. If $x_I\neq x_P$, through the time derivation, \eqref{eq:onlyAdynamics} leads to the following condition  
\begin{equation*}
\begin{aligned}
    &\frac{\dot{y}_P}{\dot{x}_P}=\frac{y_I-y_P}{x_I-x_P}\Rightarrow \frac{\ddot{y}_P\dot{x}_P-\ddot{x}_P\dot{y}_P}{\dot{x}_P^2}\\
    &=\frac{(\dot{y}_I-\dot{y}_P)(x_I-x_P)-(\dot{x}_I-\dot{x}_P)(y_I-y_P)}{(x_I-x_P)^2}.
\end{aligned}
\end{equation*}
Combining the above condition with \eqref{eq:onlyAdynamics}, we can eliminate the second-order derivatives in \eqref{eq:curvaturezero} as follows
\begin{equation}\label{eq:curvaturezero2}
\begin{aligned}
      \dod{\theta_P}{s_P}&=\dot{x}_P^2\frac{(\dot{y}_I-\dot{y}_P)(x_I-x_P)-(\dot{x}_I-\dot{x}_P)(y_I-y_P)}{v_P^3(x_I-x_P)^2}\\
      &=\frac{(\dot{y}_I-\dot{y}_P)(x_I-x_P)-(\dot{x}_I-\dot{x}_P)(y_I-y_P)}{v_P\enVert[0]{\bm{x}_I-\bm{x}_P}_2^2}\\
    &=\frac{\dot{y}_I (x_I-x_P)-\dot{x}_I (y_I-y_P)}{v_P\enVert[0]{\bm{x}_I-\bm{x}_P}_2^2}.  
\end{aligned}
\end{equation}
If $y_I\neq y_P$, we still have \eqref{eq:curvaturezero2}. According to Definition \ref{defi:interception-point}, the time derivative of $\bm{x}_I$ is
\begin{equation*}
\begin{aligned}
    \dot{x}_I &= \frac{\alpha^2 \dot{x}_E-\dot{x}_P}{ \alpha^2-1}, \\
    \dot{y}_I &= \frac{\alpha^2 \dot{y}_E-\dot{y}_P}{ \alpha^2-1} - \frac{\alpha(\bm{x}_P-\bm{x}_E)^\top(\Dot{\bm{x}}_P-\dot{\bm{x}}_E)}{(\alpha^2-1)\enVert[0]{\bm{x}_P-\bm{x}_E}_2}.
\end{aligned}
\end{equation*}
Combining the above condition with \eqref{eq:onlyAdynamics}, we can replace the terms $\dot{x}_I$ and $\dot{y}_I$ in \eqref{eq:curvaturezero2} as follows
\begin{equation*}\label{eq:thetaszero}
\begin{aligned}
    \dod{\theta_P}{s_P}& =  \frac{(\alpha^2 \dot{y}_E-\dot{y}_P)(x_I-x_P)-(\alpha^2 \dot{x}_E-\dot{x}_P)(y_I-y_P)}{v_P(\alpha^2-1)\enVert[0]{ \bm{x}_I - \bm{x}_P }_2^2} \\
    & \quad - \frac{\alpha(x_I-x_P)(\bm{x}_P-\bm{x}_E)^\top(\Dot{\bm{x}}_P-\dot{\bm{x}}_E)}{v_P(\alpha^2-1)\enVert[0]{ \bm{x}_I - \bm{x}_P }_2^2\enVert[0]{\bm{x}_P - \bm{x}_E}_2}\\
   & = \frac{F(X)^\top \bm{u}_{E}+G(X)}{\kappa}. 
\end{aligned}
\end{equation*}
Furthermore, according to \eqref{eq:pursuer_car}, the control input of the pursuer is given by
\begin{equation}\label{eq:dcontrolzeroinner}
    \begin{aligned}
        u_P&=\frac{\kappa}{v_P}\dod{\theta_P}{t}=\frac{\kappa}{v_P}\dod{s_P}{t}\dod{\theta_P}{s_P}=\kappa\dod{\theta_P}{s_P}\\
        &=F(X)^\top \bm{u}_{E}+G(X).
    \end{aligned}
\end{equation}
From the above, if $P$ adopts the feedback strategy \eqref{eq:dcontrolzeroinner}, then the positions of $P$ and $E$ will evolve along the dynamics \eqref{eq:onlyAdynamics}, thus leading to \eqref{eq:odzeroradius} for any $\bm{u}_{E}\in\mathbb{S}^1$. The only constraint for the controller \eqref{eq:dcontrolzeroinner} is $u_P\in\mathbb{S}^0$, i.e., $\envert[0]{u_P}\leq1$.

Next, we prove the necessary and sufficient condition on the capture radius $r$, minimum turning radius $\miniturnradius$ and speed ratio $\alpha$, such that $|u_P|\leq1$ holds for any $\bm{u}_{E}\in\mathbb{S}^1$, and for any $\bm{x}_P$ and $\bm{x}_E$ satisfying $\enVert[0]{\bm{x}_P-\bm{x}_E}_2\ge r$. Define 
\begin{equation*}
    x=\frac{x_P-x_E}{\enVert[0]{\bm{x}_P-\bm{x}_E}_2}, \qquad y=\frac{y_P-y_E}{\enVert[0]{\bm{x}_P-\bm{x}_E}_2}.
\end{equation*}
Note that $x$ and $y$ are two variables independent of the value of $\enVert[0]{\bm{x}_P-\bm{x}_E}_2$.
Then, according to \eqref{eq:dcontrolzeroinner}, we have 
\begin{equation*}
\begin{aligned}
\min_{(x,y)\in\mathbb{S}^1}&\min_{\bm{u}_E\in\mathbb{S}^1}u_P\\
&=\min_{(x,y)\in\mathbb{S}^1}\min_{\bm{u}_E\in\mathbb{S}^1}\big( F(X)^\top \bm{u}_{E}+G(X)\big)\\
&=\min_{(x,y)\in\mathbb{S}^1}\min_{\bm{u}_E\in\mathbb{S}^1}\Big(\frac{\kappa y(y+\alpha)u_E^x-\kappa x(y+\alpha)u_E^y}{\enVert[0]{\bm{x}_P-\bm{x}_E}_2(2\alpha y+\alpha^2+1)}\\
&\quad -\frac{\kappa\alpha x(y+\alpha)}{\enVert[0]{\bm{x}_P-\bm{x}_E}_2(2\alpha y+\alpha^2+1)^{\frac{3}{2}}}\Big)\\
&=\min_{(x,y)\in\mathbb{S}^1}\Big(\frac{-\kappa (y+\alpha)}{\enVert[0]{\bm{x}_P-\bm{x}_E}_2(2\alpha y+\alpha^2+1)}\\
&\quad -\frac{\kappa\alpha x(y+\alpha)}{\enVert[0]{\bm{x}_P-\bm{x}_E}_2(2\alpha y+\alpha^2+1)^{\frac{3}{2}}}\Big)\ge -1,
\end{aligned}
\end{equation*}
which is equivalent to 
\begin{equation*}
    \begin{aligned}
h(\alpha)&=   \max_{(x,y)\in\mathbb{S}^1}\frac{y+\alpha}{2\alpha y+\alpha^2+1}+\frac{\alpha x(y+\alpha)}{(2\alpha y+\alpha^2+1)^{\frac{3}{2}}}\\
&\leq \frac{\enVert[0]{\bm{x}_P-\bm{x}_E}_2  }{\kappa},
    \end{aligned}
\end{equation*}
holds for any $\bm{x}_P$ and $\bm{x}_E$ satisfying $\enVert[0]{\bm{x}_P-\bm{x}_E}_2\ge r$. Thus, under the condition $\enVert[0]{\bm{x}_P-\bm{x}_E}_2\ge r$, we have 
\begin{equation*}
 \min_{(x,y)\in\mathbb{S}^1}\min_{\bm{u}_E\in\mathbb{S}^1}u_P\ge-1\iff r-\kappa h(\alpha)\ge0.
\end{equation*}
Similarly, when $r-\kappa h(\alpha)\ge0$, we can obtain
\begin{equation*}
\begin{aligned}
\max_{(x,y)\in\mathbb{S}^1}&\max_{\bm{u}_E\in\mathbb{S}^1}u_P\\
&=\max_{(x,y)\in\mathbb{S}^1}\max_{\bm{u}_E\in\mathbb{S}^1}\Big(\frac{\kappa y(y+\alpha)u_E^x-\kappa x(y+\alpha)u_E^y}{\enVert[0]{\bm{x}_P-\bm{x}_E}_2(2\alpha y+\alpha^2+1)}\\
&\quad -\frac{\kappa\alpha x(y+\alpha)}{\enVert[0]{\bm{x}_P-\bm{x}_E}_2(2\alpha y+\alpha^2+1)^{\frac{3}{2}}}\Big)\\
&=\max_{(x,y)\in\mathbb{S}^1}\Big(\frac{\kappa (y+\alpha)}{\enVert[0]{\bm{x}_P-\bm{x}_E}_2(2\alpha y+\alpha^2+1)}\\
&\quad -\frac{\kappa\alpha x(y+\alpha)}{\enVert[0]{\bm{x}_P-\bm{x}_E}_2(2\alpha y+\alpha^2+1)^{\frac{3}{2}}}\Big)\\
&\leq\frac{\kappa h(\alpha)}{\enVert[0]{\bm{x}_P-\bm{x}_E}_2}\leq \frac{\kappa h(\alpha)}{r}\leq1,
\end{aligned}
\end{equation*}
which completes the proof.
\end{proof}

\begin{figure}[tbp]
    \centering
    \input{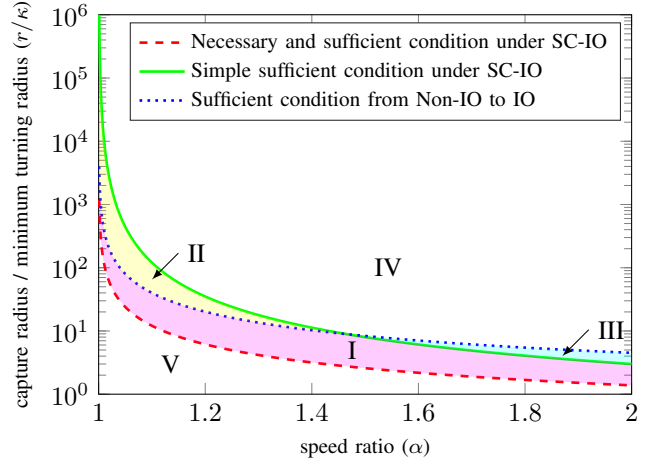}
    \caption{Summary of parameter regions for a pair of pursuer-evader.  For a pursuer $P$ and an evader $E$, if the state $X=(\bm{x}_P,\theta_P,\bm{x}_P)$ satisfies separation condition (SC) and interception orientation (IO), then the necessary and sufficient condition for $P$ winning against $E$ by the pursuit strategy \eqref{eq:dwincontrolzero} is the region above the dashed red curve, described in Theorem \ref{thom:dwinstra-car-zero}. The region above the solid green curve is a sufficient condition which is easy to be checked, described in Lemma \ref{lema:sufficient-condition-IO}. If the state $X$ does not satisfy IO (Non-IO), then a sufficient condition to steer any given Non-IO state to IO within a finite time by the strategy \eqref{eq:heading-strategy}, is the region above the dashed blue curve, described in Theorem \ref{thom:heading-adjustment}. These three curves divide the parameter space into five regions I, II, III, IV and V. Each region has different verification condition or strategy guarantee. If the state $X$ satisfies SC and Non-IO, then a sufficient condition for $P$'s winning by the two-step strategy from \eqref{eq:heading-strategy} and \eqref{eq:dwincontrolzero}, is the region above both dashed blue and red curves (II and IV), described in Theorem \ref{thom:dwin-stra-no-IO}. } 
    \label{fig:parameters}
\end{figure}

In Fig. \ref{fig:parameters}, the curve $r-\kappa h(\alpha)=0$ is plotted in red, and the condition \eqref{eq:parameterrelazero} corresponds to the region above this curve. Although \eqref{eq:parameterrelazero} is a necessary and sufficient condition such that \eqref{eq:odzeroradius} holds under the pursuit strategy \eqref{eq:dwincontrolzero}, solving \eqref{eq:alphasolutionzero} for $h(\alpha)$ might be inefficient. Next, we provide an elegant and sufficient condition to make \eqref{eq:parameterrelazero} feasible.

\begin{lema}[Sufficient condition for parameters]\label{lema:sufficient-condition-IO}
If the parameters $(r,\miniturnradius,\alpha)$ satisfy
\begin{equation}\label{eq:simpleparazero}
    \frac{r}{\kappa}\ge\frac{2\alpha-1}{(\alpha-1)^2},
\end{equation}
then $r-\kappa h(\alpha)\ge0$ in \eqref{eq:parameterrelazero} is true.   
\end{lema}
\begin{proof}
Let $\bar{h}(\alpha)$ be the optimal value of the problem
\begin{equation}\label{eq:alphasolutionzero-simple}
    \begin{aligned}
       &\underset{y\in\mathbb{R}}{\textup{maximize}}&& h_1(\alpha,y) \\
       & \textup{subject to}&& |y|\leq1,
    \end{aligned}
\end{equation}
where 
\begin{equation*}
    h_1(\alpha,y)=\frac{y+\alpha}{2\alpha y+\alpha^2+1}+\frac{\alpha (y+\alpha)}{(2\alpha y+\alpha^2+1)^{\frac{3}{2}}}.
\end{equation*}
For any $x^2+y^2=1$ and for any $\alpha>1$,
\begin{equation*}
h_1(\alpha,y)\ge \frac{y+\alpha}{2\alpha y+\alpha^2+1}+\frac{\alpha x(y+\alpha)}{(2\alpha y+\alpha^2+1)^{\frac{3}{2}}}.
\end{equation*}
Thus, the problems \eqref{eq:alphasolutionzero} and \eqref{eq:alphasolutionzero-simple} are related by $h(\alpha)\leq \bar{h}(\alpha)$. Then, we observe for $h_1(\alpha,y)$ that
\begin{equation*}
\begin{aligned}
    \dpd{h_1(\alpha,y)}{y}&=\frac{1-\alpha^2}{(2\alpha y+\alpha^2+1)^{2}}+\frac{\alpha(1-2\alpha^2-\alpha y)}{(2\alpha y+\alpha^2+1)^{\frac{5}{2}}}\\
    &<\frac{\alpha(1-\alpha^2)}{(2\alpha y+\alpha^2+1)^{\frac{5}{2}}}<0.
\end{aligned}
\end{equation*}
Thus, the optimal value $\bar{h}(\alpha)$ of \eqref{eq:alphasolutionzero-simple} is given by
\begin{equation*}
    \bar{h}(\alpha)=h_1(\alpha,-1)=\frac{2\alpha-1}{(\alpha-1)^2}.
\end{equation*}
If \eqref{eq:simpleparazero} holds, then
\begin{equation*}
\begin{aligned}
r-\kappa h(\alpha)\ge r-\kappa\bar{h}(\alpha)=r-\kappa\frac{2\alpha-1}{(\alpha-1)^2}\ge0,
\end{aligned}
\end{equation*}
which completes the proof.
\end{proof}

In Fig. \ref{fig:parameters}, the curve $\frac{r}{\kappa}=\frac{2\alpha-1}{(\alpha-1)^2}$ is plotted in green. The region above this curve, which is easy to verify, can guarantee \eqref{eq:parameterrelazero} feasible.

\begin{figure}
    \centering
    \subcaptionbox{Heading adjustment for Non-IO.}{
        \begin{tikzpicture}
        	\draw[fill, green!30] (-1.7, -0.5) rectangle (2.6, -1);
            \draw[black!50!green, very thick] (-1.8, -0.5) -- (2.7, -0.5);
        
            \node[pursuer, label=south:$\bm{x}_{P}$] (D) at (-1.5, 1) {};
        	\draw[thick, red!40, ->] (D) -- +(.55, .55) node [above, black] {$\bm{v}_P$};
        	
        	\node[point, label=right:$\bm{x}_C$] (C) at ([xshift=1cm, yshift=-1cm]D) {};
        	\draw[dashed] (D) -- node[right] {$\kappa$} (C);
        	
        	\draw[thick, red!40, densely dotted] (D) arc (135:100:1.414cm);
        	
        	\node[evader, label=right:$\bm{x}_E$] (A) at (1, 2.5) {};
            \draw[thick, blue!40, ->] (A) -- +(0.1500, -0.6330);
            
        	\draw[dashed, blue!40] (1.3125, 2.6875) circle (1.0933);
        	\node[point, label=below:$\bm{x}_I$] (I) at (1.3125, 1.5942) {};
        	
        	\draw[dashed] (D) -- (I);
        \end{tikzpicture}
    }%
    \subcaptionbox{IO is satisfied.}{%
        \begin{tikzpicture}
        	\draw[fill, green!30] (-1.5, -0.5) rectangle (2.3, -1);
            \draw[black!50!green, very thick] (-1.6, -0.5) -- (2.4, -0.5);
        
            \node[pursuer, label=north:$\bm{x}_{P}$] (D) at (0.5, 1) {};
        	\draw[thick, red!40, ->] (D) -- +(.55, -.55) node [left, black] {$\bm{v}_P$};
        	
        	\node[point, label=left:$\bm{x}_C$] (C) at ([xshift=-1cm, yshift=-1cm]D) {};
        	\draw[dashed] (D) -- node[left] {$\kappa$} (C);
        	
        	\draw[thick, red!40, densely dotted] (D) arc (45:135:1.414cm);
        	
        	\node[evader, label=right:$\bm{x}_E$] (A) at (1.39, 0.46) {};
            \draw[thick, blue!40, dotted] (1, 2.5) -- (1.39, 0.46);
            
    		\draw[dashed, blue!40] (1.5012, 0.3925) circle (-0.3904);
        	\node[point, label=below:$\bm{x}_I$] (I) at ([xshift=1cm, yshift=-1cm]D) {};
        	
        	\draw[dashed] (D) -- (I);
        \end{tikzpicture}
    }
    \caption{An example of the pursuit winning strategy for Non-IO. (a) The initial state $X$  is Non-IO. The pursuer adjusts its heading by turning clockwise (i.e., $u_P=-1$) centered at $\bm{x}_C$ with radius $\kappa$. (b) According to Theorem \ref{thom:heading-adjustment}, after a finite time, IO will be satisfied (i.e., $\theta_P=\theta_I$). Afterwards, the pursuit strategy in Theorem \ref{thom:dwinstra-car-zero} is adopted. Then, an optimal control problem is used to determine whether the pursuer can win under this two-step strategy, as stated in \thomref{thom:dwin-stra-no-IO}. The related conditions are also presented in these theorems.}
    \label{fig:heading-adjustment}
\end{figure}

\subsection{Heading Adjustment Strategy for Non-IO}\label{subsec:HAS-Non-IO}

In many cases, the state $X$ does not satisfy IO, because it is highly likely that $P$ is moving along a direction different from the interception angle $\theta_I$ when starting to capture an evader. Next, we will discuss how to design controllers for pursuer $P$ such that it can steer a given state $X$ to IO after a finite time for any evasion control input $\bm{u}_E\in\mathbb{S}^1$, as shown in Fig. \ref{fig:heading-adjustment}.

\begin{thom}[Heading adjustment strategy]\label{thom:heading-adjustment}
Consider $P\in\dteam$ and $E\in\ateam$ under the model \eqref{eq:pursuer_car} and \eqref{eq:evader_simple}, respectively. Let a Non-IO state $X$ such that $\enVert[0]{\bm{x}_P-\bm{x}_E}_2$ is sufficiently large but finite. If $P$ adopts the feedback strategy
\begin{equation}\label{eq:heading-strategy}
u_{P}=
\begin{cases}
    \mathrm{sgn}(\sin{(\theta_{I}-\theta_{P})}), & \textup{ if }\envert[0]{\theta_I-\theta_P}\neq\pi,\\
    - 1, & \textup{ otherwise,}
\end{cases}
\end{equation}
then the IO will hold after a finite time, regardless of $E$'s control input $\bm{u}_{E}\in\mathbb{S}^1$, if the parameters $(r,\miniturnradius,\alpha)$ satisfy 
\begin{equation}\label{eq:parameter-intercept-orienta}
    \frac{r}{\kappa}>\frac{(\alpha+1)^2}{\alpha(\alpha-1)}.
\end{equation}
\end{thom}

\begin{proof}
Denote by $q\in\mathbb{R}$  the inner product between the unit vector along the interception angle $\theta_I$ and the unit vector along $P$'s current heading $\theta_P$:
\begin{equation}\label{eq:inner-heading}
    q=\cos(\theta_I-\theta_P)= \frac{\Dot{\bm{x}}_P^\top (\bm{x}_I-\bm{x}_P)}{v_P \enVert[0]{\bm{x}_I-\bm{x}_P}_2}.
\end{equation}
Since $\rho( \CAR, \goal)$ is sufficiently large, according to Definition~\ref{def:interception-orientation}, the IO holds if and only if $q=1$. 

Next, we prove that the feedback strategy \eqref{eq:heading-strategy} guarantees  $\dot{q}>0$ for any $\bm{u}_{E}\in\mathbb{S}^1$ when $\enVert[0]{\bm{x}_P-\bm{x}_E}_2$ and $\rho( \CAR, \goal)$ are both sufficiently large but finite. It follows from \eqref{eq:inner-heading} that
\begin{equation}
\begin{aligned}\label{dot-inner-product}
    \dot{q} = &\frac{\ddot{\bm{x}}_P^\top(\bm{x}_I-\bm{x}_P)+\Dot{\bm{x}}_P^\top(\dot{\bm{x}}_I-\Dot{\bm{x}}_P)}{v_P \enVert[0]{\bm{x}_I-\bm{x}_P}_2}  \\
    &-\frac{\Dot{\bm{x}}_P^\top (\bm{x}_I-\bm{x}_P)}{v_P \enVert[0]{\bm{x}_I-\bm{x}_P}_2^3} (\bm{x}_I-\bm{x}_P)^\top(\dot{\bm{x}}_I-\Dot{\bm{x}}_P).
\end{aligned}
\end{equation}
Define $\bm{e}_{\theta} = [ \cos{\theta_P},\sin{\theta_P} ]^\top$ and $\ \bm{e}_{\theta}^{\perp} = [-\sin{\theta_P}, \cos{\theta_P}]^\top$. Then, according to \eqref{eq:pursuer_car} and \eqref{eq:XI}, the derivatives involved in \eqref{dot-inner-product} can be computed as follows
\begin{equation*}
\begin{aligned}
    \dot{\bm{x}}_{P}&=v_P\bm{e}_{\theta},\\   
    \ddot{\bm{x}}_P &= \frac{v_P^2 u_P}{\kappa}[-\sin{\theta_P}, \cos{\theta_P}]^\top=\frac{v_P^2 u_P}{\kappa} \bm{e}_{\theta}^{\perp},
\end{aligned}
\end{equation*}
and 
\begin{equation*}
\begin{aligned}
    \dot{\bm{x}}_I-\Dot{\bm{x}}_P
    &=  \frac{\alpha^2}{\alpha^2-1}(\dot{\bm{x}}_E-\Dot{\bm{x}}_P)  \\
    & -\frac{\alpha}{\alpha^2-1}\left[0, \frac{(\bm{x}_P-\bm{x}_E)^\top(\Dot{\bm{x}}_P-\dot{\bm{x}}_E)}{\enVert[0]{\bm{x}_P-\bm{x}_E}_2}\right]^\top\\
    &=\frac{\alpha^2}{\alpha^2-1}(v_E \bm{u}_E-v_P\bm{e}_{\theta}) \\
    &+ \frac{\alpha}{\alpha^2-1}\left[0, \frac{(\bm{x}_{P}-\bm{x}_E)^\top(v_E \bm{u}_E-v_P\bm{e}_{\theta})}{\enVert[0]{\bm{x}_{P}-\bm{x}_E}_2}\right]^\top.
\end{aligned}
\end{equation*}

Then, substituting the above conditions into \eqref{dot-inner-product} leads to
\begin{equation}
\begin{aligned}\label{eq:dot-p-2}
    \dot{q}= & \frac{1}{\enVert[0]{\bm{x}_{I}-\bm{x}_P}_2} \Big\{ \frac{v_P u_P}{\kappa}  (\bm{x}_{I}-\bm{x}_P)^\top\bm{e}_{\theta}^{\perp} + \frac{\alpha^2 v_E  \bm{u}_E^\top \bm{e}_{\theta}}{\alpha^2-1}  \\
    & - \frac{\alpha^2 v_P }{\alpha^2-1}+ \frac{\alpha \sin{\theta_P}}{ \alpha^2-1} \frac{(\bm{x}_{P}-\bm{x}_E)^\top(v_E \bm{u}_E-v_P \bm{e}_{\theta} )}{\enVert[0]{\bm{x}_{P}-\bm{x}_E}_2} \\
    & - \frac{(\bm{x}_{I}-\bm{x}_P)^\top\bm{e}_{\theta}}{\enVert[0]{ \bm{x}_{I}-\bm{x}_P}_2} \Big( \frac{\alpha^2 (\bm{x}_{I}-\bm{x}_P)^\top(v_E \bm{u}_E-v_P \bm{e}_{\theta})}{(\alpha^2-1)\enVert[0]{ \bm{x}_{I}-\bm{x}_P}_2} \\
    & + \frac{\alpha(y_I-y_P)(\bm{x}_{P}-\bm{x}_E)^\top(v_E \bm{u}_E - v_P \bm{e}_{\theta})}{(\alpha^2-1)\enVert[0]{ \bm{x}_{I}-\bm{x}_P}_2\enVert[0]{\bm{x}_{P}-\bm{x}_E}_2} \Big) \Big\}\\
    = & \frac{1}{\enVert[0]{\bm{x}_I-\bm{x}_P}_2} \Big( \frac{u_P v_P }{\kappa} \enVert[0]{ \bm{x}_I-\bm{x}_P }_2\sin{(\theta_I - \theta_P)} \\
    &+ \frac{\alpha}{\alpha^2 - 1} s(\theta_P, \theta_I, \gamma, \delta) \Big),
\end{aligned}
\end{equation}
where $\gamma\in[0,2\pi)$ and $\delta\in[0,2\pi)$ are two angles given by
\begin{equation*}
\begin{aligned}
\cos\gamma & = \frac{x_{P}-x_{E}}{\enVert[0]{\bm{x}_{P}-\bm{x}_{E}}_2},& \sin\gamma &= \frac{y_{P}-y_{E}}{\enVert[0]{\bm{x}_{P}-\bm{x}_{E}}_2},\\
\cos\delta &= u_E^x,& \sin\delta &= u_E^y,
\end{aligned}
\end{equation*}
and the part $s(\theta_P, \theta_I, \gamma, \delta)$ is as follows 
\begin{equation*}
\begin{aligned}
    &s(\theta_P, \theta_I, \gamma, \delta)\\
    &= \alpha v_E  \cos{(\theta_P - \delta)}- \alpha v_P  + v_E \sin{\theta_P} \cos{(\gamma - \delta)}\\
    &  - v_P \sin{\theta_P} \cos{(\theta_P - \gamma)}  -\alpha  v_E \cos{(\theta_I - \theta_P)} \cos{(\theta_I- \delta)}  \\
    &+ \alpha v_P \cos^2(\theta_I - \theta_P)  - v_E \sin{\theta_I} \cos{(\theta_I - \theta_P)} \cos{(\gamma - \delta)} \\
    & + v_P \sin{\theta_I} \cos{(\theta_I - \theta_P)} \cos{(\theta_P - \gamma)} \\
    &=  v_P \sin{(\theta_I-\theta_P)} (\sin{(\theta_I-\delta )} - \alpha \sin{(\theta_I-\theta_P)} \\
    & - \frac{1}{\alpha} \cos{\theta_I} \cos{(\gamma - \delta)} + \cos{\theta_I} \cos{(\gamma - \theta_P )} ),\\
    &= v_P \sin{(\theta_I-\theta_P)}\hat{s}(\theta_P, \theta_I, \gamma, \delta),
\end{aligned}
\end{equation*}
where $\hat{s}(\theta_P, \theta_I, \gamma, \delta):=\sin{(\theta_I-\delta )} - \alpha \sin{(\theta_I-\theta_P)} - \frac{1}{\alpha} \cos{\theta_I} \cos{(\gamma - \delta)} + \cos{\theta_I} \cos{(\gamma - \theta_P )}$. From the above equation,  $\hat{s}(\theta_P, \theta_I, \gamma, \delta)$ has the following bound
\begin{equation}\label{eq:s-bar-bound}
\begin{aligned}
    &\envert[0]{\hat{s}(\theta_P, \theta_I, \gamma, \delta)} \\ &= \big| \sin{(\theta_I-\delta )} - \alpha \sin{(\theta_I-\theta_P)}\\
    &\quad - \frac{1}{\alpha} \cos{\theta_I} \cos{(\gamma - \delta)} + \cos{\theta_I} \cos{(\gamma - \theta_P)} \big| \\
    & \leq (1 + \alpha +\frac{1}{\alpha} + 1) = \frac{(\alpha+1)^2}{\alpha}.
\end{aligned}
\end{equation}
There are two cases depending on if $\sin(\theta_I-\theta_P)=0$ holds.

\emph{Case 1:} Consider that $\sin(\theta_I-\theta_P)$ is nonzero. If $P$ adopts the feedback strategy \eqref{eq:heading-strategy}, it follows from \eqref{eq:dot-p-2} and \eqref{eq:s-bar-bound} that
\begin{equation}
\begin{aligned}\label{eq:dot-p-3}
    \dot{q} &\ge  \frac{v_P|\sin{(\theta_I - \theta_P)}|}{\enVert[0]{\bm{x}_I-\bm{x}_P}_2} \Big( \frac{ \enVert[0]{ \bm{x}_I-\bm{x}_P }_2}{\kappa}  \\
    &\quad\ - \frac{\alpha }{\alpha^2 - 1} \envert[0]{\hat{s}(\theta_P, \theta_I, \gamma, \delta)}  \Big)\\
    & \ge \frac{v_P|\sin{(\theta_I - \theta_P)}|}{\enVert[0]{\bm{x}_I-\bm{x}_P}_2} \Big( \frac{\enVert[0]{\bm{x}_I-\bm{x}_P }_2}{\kappa} -\frac{\alpha+1}{\alpha-1}\Big).
\end{aligned}
\end{equation}
Furthermore, for any $\bm{x}_P$ and $\bm{x}_E$ satisfying $\enVert[0]{\bm{x}_P-\bm{x}_E}_2> r$,  by \eqref{eq:XI}, $\enVert[0]{\bm{x}_I-\bm{x}_P}_2^2$ has the following bound
\begin{equation*}
\begin{aligned}
    &\enVert[0]{\bm{x}_I-\bm{x}_P}_2^2\\
    &=\frac{\alpha^4+\alpha^2}{(\alpha^2-1)^2}\enVert[0]{\bm{x}_P-\bm{x}_E}_2^2-\frac{2\alpha^3(y_E-y_P)}{(\alpha^2-1)^2}\enVert[0]{\bm{x}_P-\bm{x}_E}_2\\
    & \ge \frac{\alpha^4+\alpha^2}{(\alpha^2-1)^2}\enVert[0]{\bm{x}_P-\bm{x}_E}_2^2-\frac{2\alpha^3}{(\alpha^2-1)^2}\enVert[0]{\bm{x}_P-\bm{x}_E}_2^2\\
    & = \frac{\alpha^2}{(\alpha+1)^2}\enVert[0]{\bm{x}_P-\bm{x}_E}_2^2>\frac{\alpha^2r^2}{(\alpha+1)^2}.
\end{aligned}
\end{equation*}
Since $\enVert[0]{\bm{x}_P-\bm{x}_E}_2$ is finite, there exists a number $c\in\mathbb{R}_{>0}$ such that $\enVert[0]{\bm{x}_I-\bm{x}_P}_2\leq c$. If the parameters $(r,\miniturnradius,\alpha)$ satisfy  \eqref{eq:parameter-intercept-orienta}, then \eqref{eq:dot-p-3} implies that 
\begin{equation*}
\begin{aligned}
    \dot{q}&> \frac{v_P|\sin{(\theta_I - \theta_P)}|}{\enVert[0]{\bm{x}_I-\bm{x}_P}_2} \Big( \frac{\alpha r}{(\alpha+1)\kappa} -\frac{\alpha+1}{\alpha-1}\Big),\\
    &\ge \frac{v_P|\sin{(\theta_I - \theta_P)}|}{c} \Big( \frac{\alpha r}{(\alpha+1)\kappa} -\frac{\alpha+1}{\alpha-1}\Big)\\
    &=:c_{\text{min}}|\sin{(\theta_I - \theta_P)}|\ge0,
    \end{aligned}
\end{equation*}
holds for all $\theta_P,\theta_I,\gamma,\delta\in[0,2\pi)$, as long as $\sin(\theta_I-\theta_P)$ is nonzero and $\enVert[0]{\bm{x}_P-\bm{x}_E}_2> r$, where $c_{\text{min}}>0$.

Note that as $\theta_I-\theta_P\to0$, then $1-q\to0$ and $-\dot{q}\to0$ hold. Define $x=\theta_I-\theta_P$ and $V(x)=1-q=1-\cos(x)$. Then, for any  $x\in(-2/\pi,2/\pi)$, we have
\begin{equation*}
\begin{aligned}
\dot{V}(x)=-\dot{q}\leq -c_{\text{min}}\envert[0]{\sin(x)}\leq -\frac{2c_{\text{min}}}{\pi}\envert[0]{x}\leq -\frac{2c_{\text{min}}}{\pi}\sqrt{V(x)}.
\end{aligned}
\end{equation*}
It follows from \cite[Theorem 4.2]{SPB-DSB:00} that $x=0$ is a finite-time-stable equilibrium. Therefore, $q=1$ can be reached within a finite time.

\emph{Case 2:} If $\sin(\theta_I-\theta_P)=0$, then $\envert[0]{\theta_I-\theta_P}=\pi$ holds. Note that \eqref{eq:dot-p-2} implies that $\dot{q}=0$. Furthermore, we take $u_P=-1$ as in \eqref{eq:heading-strategy}. According to \eqref{eq:dot-p-2} and the bound \eqref{eq:s-bar-bound}, the second-order time derivative of $q$ at this point is
\begin{equation*}
\begin{aligned}
  \eval[1]{\ddot{q}}_{\envert[0]{\theta_I-\theta_P}=\pi}&=  \frac{-1}{\enVert[0]{\bm{x}_I-\bm{x}_P}_2} \Big( \frac{u_P v_P }{\kappa} \enVert[0]{ \bm{x}_I-\bm{x}_P }_2 \\
    &\quad + \frac{\alpha v_P}{\alpha^2 - 1} \hat{s}(\theta_P, \theta_I, \gamma, \delta) \Big),\\
    &\ge\frac{v_P}{\enVert[0]{\bm{x}_I-\bm{x}_P}_2}\Big( \frac{\enVert[0]{\bm{x}_I-\bm{x}_P }_2}{\kappa} -\frac{\alpha+1}{\alpha-1}\Big)\\
    &>\frac{v_P}{\enVert[0]{\bm{x}_I-\bm{x}_P}_2} \Big( \frac{\alpha r}{(\alpha+1)\kappa} -\frac{\alpha+1}{\alpha-1}\Big)\ge0.
\end{aligned}
\end{equation*}
Therefore, $X$ will deviate from $\envert[0]{\theta_I-\theta_P}=\pi$ as time goes by, combing back to Case 1. Thus, we complete the proof.
\end{proof}

\thomref{thom:heading-adjustment} shows that if $P$ and $E$ are far from each other initially, then the heading adjustment strategy \eqref{eq:heading-strategy} can steer a Non-IO state $X$ to achieve IO within a finite time, no matter what strategy the evader adopts. Thus, the Non-IO case becomes the IO case for which we have proposed a pursuit winning strategy. In Fig. \ref{fig:parameters}, the curve $\frac{r}{\kappa}=\frac{(\alpha+1)^2}{\alpha(\alpha-1)}$ is plotted in blue, and the region above this curve can sufficiently guarantee the transfer from the Non-IO to the IO via strategy \eqref{eq:heading-strategy} within a finite time.


\subsection{Pursuit Winning Strategy for SC and Non-IO}
Now, we will present the pursuit winning strategy when the state $X$ satisfies SC but does not satisfy IO, by merging the strategies proposed in Sections \ref{subsec:PWS-IO} and \ref{subsec:HAS-Non-IO}.

Suppose that the pursuer $P$ adopts the strategy \eqref{eq:heading-strategy} aiming to reach IO, while the evader $E$ strives to put the interception point when $q=1$ (i.e., the IO holds), at a position having a minimum signed distance to $\targetline$, because once the IO is achieved, the distance between the interception point and $\targetline$ decides the game winner (i.e., whether the SC holds), as \thomref{thom:dwinstra-car-zero} shows. Formally, the terminal set $\Psi$ and payoff function $J$ respectively are
\begin{equation}
\begin{aligned}\label{eq:optimal-control-twosets}
\Psi =\Psi_1\cup\Psi_2,\quad  J=\rho_{\targetline}(\bm{x}_I(t_f)),
\end{aligned}
\end{equation}
where $\Psi_1=\{X\,|\,\theta_{P}=\theta_{I}\}$, $\Psi_2=\{X\,|\,\enVert[0]{\bm{x}_P-\bm{x}_E}_2\leq r\}$, and the terminal time $t_f\in\mathbb{R}_{\ge0}$ is defined as the time instant when the state $X$ enters $\Psi$. Suppose that \eqref{eq:parameter-intercept-orienta} holds. Then, $t_f$ is finite, because by \thomref{thom:heading-adjustment}, if $P$ and $E$ are close to each other, $E$ will be captured before achieving IO.

\begin{lema}[Optimal evasion control]\label{lema:straight-attack-control} Consider the optimal control problem \eqref{eq:pursuer_car}, \eqref{eq:evader_simple}, and \eqref{eq:optimal-control-twosets}. The optimal control of the evader $E$ is constant and the optimal evasion trajectory is a straight line.  
\end{lema}
\begin{proof}
The Hamiltonian associated with this problem is
\begin{equation*}
\begin{aligned}
     \mathcal{H}=\lambda_1 v_{P}\cos\theta_P&+\lambda_2v_P\sin\theta_P+\lambda_3v_Pu_P/\kappa\\
     &+\lambda_4 v_E u_E^x+\lambda_5 v_E u_E^y,  
\end{aligned}
\end{equation*}
where $\lambda_i\in\mathbb{R}$ $(i=1,2,\dots,5)$ is the Lagrange multiplier, and $u_P=1$ or $-1$. The optimal control $\bm{u}_E^*=[u_E^{x*},u_E^{y*}]^\top$ satisfies
\begin{equation*}
\begin{aligned}
\bm{u}_E^*=    \underset{\bm{u}_E\in\mathbb{S}^1}{\textup{argmin}} \ \mathcal{H}=\left[\frac{-\lambda_4}{\sqrt{\lambda_4^2+\lambda_5^2}},\frac{-\lambda_5}{\sqrt{\lambda_4^2+\lambda_5^2}}\right]^\top.
\end{aligned}
\end{equation*}
The costate equations with respect to $\lambda_4$ and $\lambda_5$ are
\begin{equation*}
\begin{aligned}
    \dot{\lambda}_4=\pd{\mathcal{H}}{{x_E}}=0,\quad \dot{\lambda}_5=\pd{\mathcal{H}}{{y_E}}=0.
\end{aligned}
\end{equation*}
Thus, $\bm{u}_E^*$ is constant, as $\lambda_4$ and $\lambda_5$ are both constant. Furthermore, the optimal trajectory of $E$ is a straight line, which completes the proof.
\end{proof}


Since the optimal control of the evader $E$ is constant, the optimal control problem \eqref{eq:pursuer_car}, \eqref{eq:evader_simple}, and \eqref{eq:optimal-control-twosets} can be reformulated as the following optimization problem.

\begin{pbm}[Minimal signed distance to $\targetline$]\label{prob:lowest-IP}
For any state $X(t)=(\bm{x}_{P},\theta_{P},\bm{x}_{E})$, define $\rho^*_{\targetline}(X)$ be the optimal value of the problem
\begin{subequations}\label{eq:op-prob}
    \begin{align}
       &\textup{minimize}&& \rho_{\targetline}(\bm{x}_{\tau}(\theta_E)) \\
       & \textup{variables}&& \theta_E\in[0,2\pi),\,
       \tau\in[t,+\infty)\\
       & \textup{subject to}&& X_{\tau}(\theta_E)\in\Psi,\label{subeq:pbm1-constraint-final}\\
       &&& X_s(\theta_E)\notin\Psi,\quad \forall\, t\leq s< \tau, \label{subeq:pbm1-constraint-interval}
    \end{align}
\end{subequations}
where for every $s\in[t,\tau]$,
\begin{subequations}
    \begin{align}
    & \bm{x}_E^{+}=\bm{x}_E+v_E(s-t)[\cos\theta_E,\sin\theta_E]^\top,\label{subeq:pbm1-x-A-plus}\\
      &\begin{aligned}
      \hspace{-0.02cm}\bm{x}_P^{+}&=\bm{x}_P+\textup{sgn}(\sin(\theta_I-\theta_P))\kappa[\sin\theta_P^+,-\cos\theta_P^+]^\top\label{subeq:pbm1-x-D-plus}\\
      &\qquad\ \ \hspace{0.02cm} -\textup{sgn}(\sin(\theta_I-\theta_P))\kappa[\sin\theta_P,-\cos\theta_P],
      \end{aligned} \\
      & \hspace{0.05cm}\theta_P^+=\theta_P+\frac{v_P(s-t)\textup{sgn}(\sin(\theta_I-\theta_P))}{\kappa},\label{subeq:pbm1-theta-D-plus}\\
      &\hspace{0.04cm}\Upsilon^+=\Big[
    0, - \frac{\alpha \enVert[0]{\bm{x}_P^+ - \bm{x}_E^+ }_2}{\alpha^2-1}
\Big]^\top,\label{subeq:pbm1-d-plus}\\
      & \hspace{0.04cm}\bm{x}_{s}(\theta_E)=\frac{\alpha^2 \bm{x}_E^+-\bm{x}_P^+}{\alpha^2-1}+\Upsilon^+,\label{subeq:pbm1-x-t}\\
    & X_s(\theta_E)=\big(\bm{x}^+_P,\theta_P^+,\bm{x}_E^+\big).\label{subeq:pbm1-X-t}
    \end{align}
\end{subequations}
If the problem is infeasible, we define $\rho^*_{\targetline}(X)=+\infty$.
\end{pbm}

Since $P$ adopts the strategy \eqref{eq:heading-strategy}, then it moves along  \eqref{subeq:pbm1-x-D-plus} and \eqref{subeq:pbm1-theta-D-plus}, where $(\bm{x}_P^+,\theta_P^+)$ is the state of $P$ at time $s$ starting from $(\bm{x}_P,\theta_P)$ at time $t$. Suppose that $E$ follows the strategy corresponding to the solution to the optimal control problem \eqref{eq:pursuer_car}, \eqref{eq:evader_simple}, and \eqref{eq:optimal-control-twosets}. According to Lemma \ref{lema:straight-attack-control}, the optimal control is constant. Thus, $E$ moves along \eqref{subeq:pbm1-x-A-plus} if $\theta_E$ is the optimal heading angle, where $\bm{x}_E^+$ is the state of $E$ at time $s$ starting from $\bm{x}_E$ at time $t$. Note that \eqref{subeq:pbm1-d-plus}-\eqref{subeq:pbm1-X-t} are the computations of $\bm{x}_s(\theta_E)$ and $X_s(\theta_E)$ by $\bm{x}^+_P$, $\theta_P^+$ and $\bm{x}_E^+$. If the system state hits $\Psi_1$ before $\Psi_2$, then $\tau^*$ is $P$'s heading adjustment time.

Next, we relate the optimal value to Problem \ref{prob:lowest-IP} with the outcomes when $P$ adopts the heading adjustment strategy.

\begin{lema}[Function of the optimal value]\label{lema:func-optimal-value} For any state $X$, let $(\theta_E^*,\tau^*)$ be an optimal solution to Problem \ref{prob:lowest-IP} when $\rho^*_{\targetline}(X)<+\infty$. If \eqref{eq:parameter-intercept-orienta} holds and $P$ adopts the strategy \eqref{eq:heading-strategy}, then
\begin{enumerate}[label=(\roman*)]
    \item\label{itm:less-zero} if $\rho^*_{\targetline}(X)<0$, then by strategy $\bm{u}_E=[\cos\theta_E^*,\sin\theta_E^*]^\top$, there exists a time instant $s\in[t,\tau^*]$ at which the SC fails before achieving IO or the capture;\label{itm:optimal-value-negative}
    \item\label{itm:finite-positive} if $0\leq\rho^*_{\targetline}(X)<+\infty$, then there exists a finite time instant $s\in[t,+\infty)$ at which the SC holds, and furthermore IO holds or $E$ is captured, regardless of $E$'s strategy;\label{itm:optimal-value-positive}
    \item\label{itm:infinite-positive} if $\rho^*_{\targetline}(X)=+\infty$, then $X\in \Psi$, that is, at time $t$, the IO holds or $E$ is captured by $P$. \label{itm:optimal-value-infinite}
\end{enumerate}
\end{lema}
\begin{proof}
The constraints \eqref{subeq:pbm1-constraint-final} and \eqref{subeq:pbm1-constraint-interval} are feasible if and only if $
X\notin\Psi$, i.e., $X_t(\theta_E)\notin\Psi$, because the strategy \eqref{eq:heading-strategy} can always steer $X_{\tau}(\theta_E)$ into $\Psi_1$ within a finite time, even if $X_{\tau}(\theta_E)$ does not enter $\Psi_2$. Thus, if $\rho^*_{\targetline}(X)=+\infty$, we have $X\in\Psi$, implying that \ref{itm:optimal-value-infinite} is straightforward.

Regarding \ref{itm:optimal-value-negative}, $\rho^*_{\targetline}(X)<0$ implies that if $E$ adopts the strategy $\bm{u}_E=[\cos\theta_E^*,\sin\theta_E^*]^\top$, then it can steer $\bm{x}_s(\theta_E)$ into $\goal$ before $X_s(\theta_E)$ getting into $\Psi$. Thus, the conclusion is obtained. 

Regarding \ref{itm:optimal-value-positive}, $\rho^*_{\targetline}(X)<+\infty$ implies that no matter what strategy $E$ adopts, $P$ can steer $X_s(\theta_E)$ into $\Psi$ at a time instant $s\in[t,\tau^*]$, i.e., either $E$ is captured or the IO is satisfied. Additionally, $\rho^*_{\targetline}(X)\ge0$ means that the SC still holds at the time instant $s$.  
\end{proof}

Although the optimal value to Problem \ref{prob:lowest-IP} is attractive, it is hard to compute it directly due to the infinite constraints and the non-convexity. We next present an approximation solution to Problem \ref{prob:lowest-IP}, thus providing a sufficient condition for the pursuer's winning. We first give a nontrivial upper bound of the heading adjustment time.

\begin{figure}[tb]
    \centering
    \begin{tikzpicture}
        \coordinate[label = left:$\bm{x}_C$] (O) at (0,0);
        \fill (O) circle (0.05);
        \draw ({5/sqrt(26)}, {1/sqrt(26)}) arc (11.30993:135:1);
        \draw[dashed] (5, 1) circle ({sqrt(2)});
        \coordinate[label = right: $\bm{x}_E$] (A) at (5, 1);
        \coordinate[label = above left: $\bm{x}_E^+$] (B) at (4, 2);
        \node [evader] at (5, 1) {};
        \node [evader] at (4, 2) {};
        \draw[->] (A) -- (B);
        \node [pursuer] at ({-sqrt(2)/2}, {sqrt(2)/2}) {};
        \coordinate[label = above left: $\bm{x}_P$] (C) at ({-sqrt(2)/2}, {sqrt(2)/2});
        \draw[dashed] (C) arc (135:371.30993:1);
        \coordinate[label = above: $\theta_P$] (F) at ({-sqrt(2)/2+0.5}, {sqrt(2)/2+0.5});
        \draw[->] (C) -- (F);
        \draw[dashed] (0, 0) -- (A);
        \coordinate (D) at ({5/sqrt(26)}, {1/sqrt(26)});
        \coordinate[label=above right: $\bm{x}_A$] (A-P) at ({5/sqrt(26)-0.15}, {1/sqrt(26)+0.1});
        \coordinate (E) at ({4/sqrt(20)}, {2/sqrt(20)});
        \fill (D) circle (0.05);
        \coordinate[label = right: $x$] (G) at ({5/sqrt(26) + 1/sqrt(13)}, {1/sqrt(26) - 5/sqrt(13)});
        \coordinate[label = below: $y$] (H) at ({5/sqrt(26) + 5/sqrt(13)}, {1/sqrt(26) + 1/sqrt(13)});
        \coordinate[label = below: $\mathbb{C}_1$] (C1) at (-1.2,-0.2);
         \coordinate[label = below: $\kappa$] (KA) at (0.4,0.5);
        \coordinate[label = below: $\mathbb{C}_2$] (C2) at (5.3,-0.4);
         \coordinate[label = below: $v_E\Delta$] (ED) at (4.9,2);
        \draw[->] (D) -- (G);
        \draw[->] (D) -- (H);
    \end{tikzpicture}
    \caption{An upper bound of the heading adjustment time such that the IO is satisfied, when the initial state $(\bm{x}_P,\theta_P,\bm{x}_E)$ is Non-IO. The pursuer $P$ at $\bm{x}_P$ swerves along the circle $\mathbb{C}_1$ with the minimum turning radius $\kappa$ and center $\bm{x}_C$, toward the direction given by the heading adjustment strategy \eqref{eq:heading-strategy}. It is proved that under certain conditions, the IO will be satisfied before $P$ reaches $\bm{x}_A$, where $\bm{x}_A$ is the point on $\mathbb{C}_1$ closest to $\bm{x}_E$. The circle $\mathbb{C}_2$ is the set of $E$'s possible positions under maximal constant evasion controls when $P$ reaches $\bm{x}_A$.}
    \label{fig:heading_adjustmentl}
\end{figure}
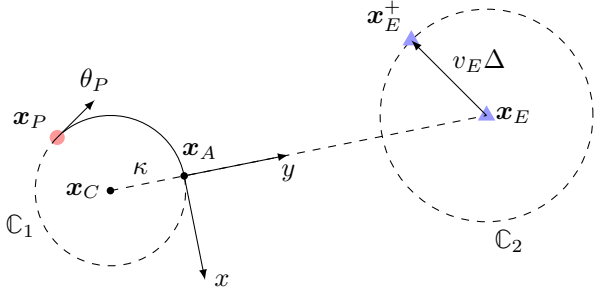

\begin{lema}[Heading adjustment time, upper bound]\label{lema:Heading-modi-time} 
Let a state $X=(\bm{x}_{P},\theta_{P},\bm{x}_{E})$ satisfying the following conditions:
\begin{enumerate}[label=(\roman*)]
    \item\label{itm:no-capture} $\enVert[0]{\bm{x}_P-\bm{x}_E}_2>r$;
    \item\label{itm:no-IO} $\theta_P\neq \theta_I$;
    \item\label{item:no-close} $\enVert[0]{\bm{x}_C-\bm{x}_E}_2>\kappa+v_P\Delta(X)/\sqrt{\alpha^2-1}$. The variables satisfy $\bm{x}_C=[x_C,y_C]^\top=\bm{x}_P+\kappa[\cos\theta_C,\sin\theta_C]^\top$, $\theta_C=\theta_P+\frac{\pi}{2}\textup{sgn}(\sin(\theta_I-\theta_P))$, and
\begin{equation}
   \Delta(X)=\frac{\left[(\theta_E-\theta_C-\pi)\textup{sgn}(\sin(\theta_I-\theta_P))+2\pi n\right]\kappa}{v_P},
\end{equation}
where $\theta_E\in[0,2\pi)$ is given by $[\cos\theta_E,\sin\theta_E]^\top=\frac{\bm{x}_E-\bm{x}_C}{\enVert[0]{\bm{x}_E-\bm{x}_C}_2}$ and $n$ is the unique non-negative integer such that $\Delta(X)\in(0,2\pi\kappa/v_P]$ holds.
\end{enumerate}
If \eqref{eq:parameter-intercept-orienta} holds, for any optimal solution $(\theta_E^*,\tau^*)$ to Problem \ref{prob:lowest-IP}, we have $ t\leq \tau^*\leq t+\Delta(X)$.
\end{lema}
\begin{proof}
Note that $X$ satisfies conditions \ref{itm:no-capture} and \ref{itm:no-IO}, so the value of $\textup{sgn}(\theta_I-\theta_P)$ can be determined. Without loss of generality, let $\textup{sgn}(\theta_I-\theta_P)=-1$, i.e., $P$ turns in the clockwise direction as Fig. \ref{fig:heading_adjustmentl} shows. Thus, $P$ moves along a circular orbit $\mathbb{C}_1=\{\bm{x} \in \mathbb{R}^2\, |\, \enVert[0]{\bm{x} - \bm{x}_C}_2 = \kappa\}$ where $\bm{x}_C=\bm{x}_P+\kappa[\cos(\theta_P-\frac{\pi}{2}),\sin(\theta_P-\frac{\pi}{2})]^\top$. 

According to the condition \ref{itm:infinite-positive} in Lemma \ref{lema:func-optimal-value}, Problem \ref{prob:lowest-IP} is feasible, becuase $X\notin\Psi$ under the conditions \ref{itm:no-capture} and \ref{itm:no-IO}. Let $(\theta_E^*,\tau^*)\in[0,2\pi)\times[t,+\infty)$ be any optimal solution to Problem \ref{prob:lowest-IP}.

Let $\bm{x}_A$ be the projection of $\bm{x}_E$ on the orbit $\mathbb{C}_1$ in Fig. \ref{fig:heading_adjustmentl}. For $P$, the time of moving from $\bm{x}_P$ to $\bm{x}_A$ along $\mathbb{C}_1$ is
\begin{equation*}\label{eq:time-te}
    \Delta(X)=\frac{\left[(\theta_E-\theta_C-\pi)\textup{sgn}(\sin(\theta_I-\theta_P))+2\pi n\right]\kappa}{v_P}>0,
\end{equation*}
where $\theta_E\in[0,2\pi)$ is the angle of $\bm{x}_E-\bm{x}_C$ and $n$ is the unique non-negative integer such that $\Delta\in[0,2\pi\kappa/v_P]$ holds. Additionally, $\Delta>0$ is because $X\notin\Psi$ and $P$ turns in the clockwise direction. When $P$ reaches $\bm{x}_A$, $E$'s extreme position could be any point on the circle $\mathbb{C}_2 = \{\bm{x} \in \mathbb{R}^2\, |\, \enVert[0]{\bm{x} - \bm{x}_E}_2 = v_E \Delta\}$, depending on the value of $\theta_E$. Let $\bm{x}_E^+=[x_E^+,y_E^+]\in\mathbb{C}_2$ and $\bm{x}_P^+=\bm{x}_A$ be $E$'s and $P$'s positions at time $t+\Delta$, respectively. Thus, the closure of the ER at time $t+\Delta$ is
\begin{equation*}
\begin{aligned}
&\CAR(t+\Delta)\\
&=\Big\{\bm{x}\in\mathbb{R}^2\,\big|\,\enVert[2]{\bm{x}-\frac{\alpha^2\bm{x}^+_E-\bm{x}_P^+}{\alpha^2-1}}_2 \leq \frac{\alpha\enVert[0]{\bm{x}_P^+-\bm{x}_E^+}_2}{\alpha^2-1}\Big\}.
\end{aligned}
\end{equation*}

 Note that the condition \ref{item:no-close} implies that $\enVert[0]{\bm{x}_C-\bm{x}_E}_2>\kappa$, i.e., $\bm{x}_E$ lies out of $\mathbb{C}_1$. For simplicity of description, we build a local Cartesian coordinate system with $\bm{x}_A$ as the origin, the tangent line of $\mathbb{C}_1$ at $\bm{x}_A$ as the $x$-axis, and $\bm{x}_E-\bm{x}_A$ as the $y$-axis. In the local coordinate system, we have $\bm{x}_P^+=\bm{0}$ and $\bm{x}_E=[0,y_E]^\top$ with $y_E>0$. Thus, for any given $\bm{x}_E^+\in\mathbb{C}_2$, the minimal $y$-coordinate of points on $\CAR(t+\Delta)$ is
 \begin{equation}
 \begin{aligned}\label{eq:min-y-t+tE}
  &\min_{\bm{x}\in\CAR(t+\Delta)}\ y\\
  &=\frac{\alpha^2y_E^+ -y_P^+}{\alpha^2-1}- \frac{\alpha\enVert[0]{\bm{x}_P^+-\bm{x}_E^+}_2}{\alpha^2-1}\\
  &=\frac{\alpha^2}{\alpha^2-1}y_E^+ - \frac{\alpha\enVert[0]{\bm{x}_E^+}_2}{\alpha^2-1}\\
  &= \frac{\alpha^2}{\alpha^2-1}y_E^+ - \frac{\alpha}{\alpha^2-1}\sqrt{v_E^2\Delta^2+y_E(2y_E^+-y_E)}.
 \end{aligned}
 \end{equation}
We consider a function $h_2(y_E^+)=\alpha^2(y_E^{+})^2-2y_Ey_E^++y_E^2-v_E^2\Delta^2$ in $y_E^+$. Then, we have 
\begin{equation}\label{eq:y-E-h-2}
\min_{\bm{x}\in\CAR(t+\Delta)}\ y>0 \iff y_E^+\ge0,\ h_2(y_E^+)>0.
\end{equation}

Next we analyze the minimal value of $h_2(y_E^+)$. Since $\bm{x}_E^+\in\mathbb{C}_2$, then $y_E^+\in[y_E-v_E\Delta,y_E+v_E\Delta]$. Thus, if $(\alpha^2-1)y_E\ge \alpha^2v_E\Delta$, then
\begin{equation}
\begin{aligned}\label{eq:h3-1}
\min_{\bm{x}_E^+\in\mathbb{C}_2}h_2(y_E^+)&=h_2(y_E-v_E\Delta)\\
&=(\alpha^2-1)(y_E-v_E\Delta)^2\ge\frac{v_E^2\Delta^2}{\alpha^2-1}.
\end{aligned}
\end{equation}
If $(\alpha^2-1)y_E< \alpha^2v_E\Delta$, then
\begin{equation}
\begin{aligned}\label{eq:h3-2}
\min_{\bm{x}_E^+\in\mathbb{C}_2}h_2(y_E^+)=h_2\big(\frac{1}{\alpha^2}y_E\big)=\frac{(\alpha^2-1)y_E^2}{\alpha^2}-v_E^2\Delta^2.
\end{aligned}
\end{equation}
By combining \eqref{eq:h3-1} and \eqref{eq:h3-2}, we have
\begin{equation}\label{eq:min-h3}
   \min_{\bm{x}_E^+\in\mathbb{C}_2}h_2(y_E^+)>0\iff y_E>\frac{\alpha v_E\Delta}{\sqrt{\alpha^2-1}}.
\end{equation}
Since $y_E^+\in[y_E-v_E\Delta,y_E+v_E\Delta]$, we have
\begin{equation}\label{eq:y-E-plus-inequ}
\min_{\bm{x}_E^+\in\mathbb{C}_2}y_E^+\ge0\iff y_E\ge v_E\Delta.
\end{equation}
Thus, by \eqref{eq:min-y-t+tE}, \eqref{eq:y-E-h-2}, \eqref{eq:min-h3} and \eqref{eq:y-E-plus-inequ}, the minimal $y$-coordinate of points on $\CAR(t+\Delta)$ for all $\bm{x}_E^+\in\mathbb{C}_2$, is positive, i.e.,
\begin{equation}\label{eq:min-min-y}
  \min_{\bm{x}_E^+\in\mathbb{C}_2}\min_{\bm{x}\in\CAR(t+\Delta)}\ y >0,
\end{equation}
if and only if
\begin{equation*}
\begin{aligned}
\|\bm{x}_C-\bm{x}_E\|_2&= \kappa + y_E> \kappa+ \frac{\alpha v_E\Delta}{\sqrt{\alpha^2-1}},
\end{aligned}
\end{equation*}
that is, the condition \ref{item:no-close} holds.

Let $\theta_P^+\in[0,2\pi)$ and $\theta_I^+\in[0,2\pi)$ be $P$'s heading and interception angle at time $t+\Delta$, respectively. Note that \eqref{eq:min-min-y} implies that for any $\theta_E\in[0,2\pi)$, the interception point $\bm{x}_I$ at time $t+\Delta$ lies at the left side of $\theta_P^+$. Thus, $\textup{sgn}(\theta_P^+-\theta_I^+)=1$, that is, $P$ needs to turn in the counterclockwise direction. According to the continuity of $\theta_I-\theta_P$ in time $t$, there exists a time $s\in[t,t+\Delta]$ such that $\theta_P=\theta_I$ at time $s$. Thus, for any optimal solution $(\theta_E^*,\tau^*)\in[0,2\pi)\times[t,+\infty)$, we have $t\leq \tau^*\leq t+\Delta$, which completes the proof.
\end{proof}

According to \lemaref{lema:Heading-modi-time}, if the pursuer $P$ swerves with the minimum turning radius and toward the direction given by the heading adjustment strategy \eqref{eq:heading-strategy}, then $P$ will achieve IO or capture $E$ before combing back to its starting position. In other words, $P$ needs at most one period to achieve IO or capture $E$. Based on this characteristic, we next introduce a simpler optimization problem, which is a relaxation of Problem \ref{prob:lowest-IP}.

\begin{pbm}[Minimal signed distance to $\targetline$, lower bound]\label{prob:lowest-IP-lb}
For any state $X=(\bm{x}_{P},\theta_{P},\bm{x}_{E})$, define $\widehat{\rho}_{\targetline}(X)$ be the optimal value of the problem
\begin{subequations}\label{eq:op-prob-lb}
    \begin{align}
       &\textup{minimize}&& (\alpha^2-1)\rho_{\targetline}(\bm{x}_{I}^+) \label{subeq:pbm2-obj-func}\\
       & \textup{variables}&& \bm{x}_P^+\in\mathbb{R}^2,\bm{x}_E^+\in\mathbb{R}^2\\
       & \textup{subject to}&& 
       \enVert[0]{\bm{x}_E^+-\bm{x}_E}_2\leq \frac{2\pi\kappa}{\alpha},\label{subeq:pbm2-constraint-1}\\
       &&&\hspace{-0.02cm}\enVert[0]{\bm{x}_P^{+}-\bm{x}_C}_2\leq \kappa,\label{subeq:pbm2-constraint-2}
       \\
       &&&\Upsilon^+=\Big[
    0, - \frac{\alpha\enVert[0]{ \bm{x}_P^+ - \bm{x}_E^+ }_2}{\alpha^2-1}
    \Big]^\top,\\
      &&& \hspace{0.04cm}\bm{x}_{I}^+=\frac{\alpha^2 \bm{x}_E^+-\bm{x}_P^+}{\alpha^2-1}+\Upsilon^+.
\end{align}
\end{subequations}
\end{pbm}

\begin{lema}[Problem \ref{prob:lowest-IP-lb} is a relaxation]\label{lema:pbm2-relaxation} Let a state $X$ such that the conditions \ref{itm:no-capture}-\ref{item:no-close} in Lemma \ref{lema:Heading-modi-time} are true. Then
\begin{enumerate}[label=(\roman*)]
    \item\label{itm:nonconvex-pbm2-1} the optimal values to Problems \ref{prob:lowest-IP} and \ref{prob:lowest-IP-lb} are related by $\rho_{\targetline}^*(X)\ge\widehat{\rho}_{\targetline}(X)$;
    \item \label{itm:nonconvex-pbm2-2} Problem \ref{prob:lowest-IP-lb} is a non-convex optimization problem with the convex constraint set and concave objective function;
       \item\label{itm:nonconvex-pbm2-3} suppose that \eqref{eq:parameter-intercept-orienta} holds and $P$ adopts the strategy \eqref{eq:heading-strategy}. If $\widehat{\rho}_{\targetline}(X)\ge0$, then there exists a finite time instant $s\in[t,+\infty)$ at which the SC holds, and furthermore IO holds or $E$ is captured, regardless of $E$'s strategy.
\end{enumerate}
\end{lema}
\begin{proof}
Regarding \ref{itm:nonconvex-pbm2-1}, \lemaref{lema:Heading-modi-time} has proved that $P$ needs at most one period to achieve IO or capture $E$ along $\mathbb{C}_1$. Thus, the constraint in  \pbmref{prob:lowest-IP-lb} allows two larger sets of reachable positions for $P$ and $E$ than \pbmref{prob:lowest-IP}. The difference of objective functions between Problems \ref{prob:lowest-IP} and \ref{prob:lowest-IP-lb} is a positive constant. Thus, the conclusion is obtained.

Regarding \ref{itm:nonconvex-pbm2-2}, note that \eqref{subeq:pbm2-constraint-1} and \eqref{subeq:pbm2-constraint-2} are independent and both convex. Thus, the constraint set is convex. The objective function \eqref{subeq:pbm2-obj-func} can be rewritten as
\begin{equation*}
  (\alpha^2-1)\rho_{\targetline}(\bm{x}_{I}^+) = \alpha^2y_E^+-y_P^+-\alpha\enVert[0]{\bm{x}_P^+-\bm{x}_E^+}_2,
\end{equation*}
which is concave on the interval $(\bm{x}_P^+,\bm{x}_E^+)\in\mathbb{R^2}\times\mathbb{R}^2$. Thus, Problem \ref{prob:lowest-IP-lb} is a non-convex. 

Regarding \ref{itm:nonconvex-pbm2-3}, it directly follows from \ref{itm:nonconvex-pbm2-1} and the conclusion \ref{itm:finite-positive} in \lemaref{lema:func-optimal-value}.
\end{proof}

Next, the non-convex Problem \ref{prob:lowest-IP-lb} is addressed by a sextic equation which can be solved more efficiently.

\begin{thom}[Solution to Problem \ref{prob:lowest-IP-lb}]\label{thom:solution-pbm-2}
Let a state $X=(\bm{x}_{P},\theta_{P},\bm{x}_{E})$ such that the conditions \ref{itm:no-capture}-\ref{item:no-close} in Lemma \ref{lema:Heading-modi-time} are true. Let $\sigma=\textup{sgn}(x_C-x_E)$. Then
\begin{enumerate}[label=(\roman*)]
    \item \label{itm:solution-sign-pbm2} if $(\bm{x}_P^*,\bm{x}^*_E)$ is an optimal solution to Problem \ref{prob:lowest-IP-lb}, then $\textup{sgn}(x^*_P-x_C)=\textup{sgn}(x_E-x_E^*)=\textup{sgn}(x_P^\ast-x_E^\ast)=\sigma$;
    \item \label{itm:solution-pbm2} the optimal value to Problem \ref{prob:lowest-IP-lb} is
    \begin{equation*}
        \widehat{\rho}_{\targetline}(X)=\alpha^2y_E^*-y_P^*-\alpha\enVert[0]{\bm{x}_P^*-\bm{x}_E^*}_2,
    \end{equation*}
     and the optimal solution $(\bm{x}_P^*,\bm{x}^*_E)$ is
    \begin{equation*}
    \begin{aligned}
    x_P^*&=x_C+\frac{\sigma\kappa}{2\lambda}\sqrt{2(1+\alpha^2)\lambda^2-\lambda^4-(1-\alpha^2)^2},\\
    y_P^*&=y_C+\frac{(1-\alpha^2+\lambda^2)\kappa}{2\lambda},\\
    x_E^*&=x_E-\frac{\sigma\pi\kappa}{\alpha^2\lambda}\sqrt{2(1+\alpha^2)\lambda^2-\lambda^4-(1-\alpha^2)^2},\\
    y_E^*&=y_E+\frac{(1-\alpha^2-\lambda^2)\pi\kappa}{\alpha^2\lambda},
    \end{aligned}
    \end{equation*}
    where $\lambda\in\mathbb{R}_{>0}$ is the solution of the sextic equation
    \begin{equation}\label{eq:lambda-six}
        k_6\lambda^6+k_5\lambda^5+k_4\lambda^4+k_3\lambda^3+k_2\lambda^2+k_1\lambda^1+k_0=0,
    \end{equation}
    which is parameterized by
    \begin{equation*}
    \begin{aligned}
    &k_6=\enVert[0]{\bm{x}_C-\bm{x}_E}_2^2,\ k_5=\kappa(4\pi+2)(y_C-y_E),\\
    &k_4=(2\pi+1)^2\kappa^2-2(1+\alpha^2)\enVert[0]{\bm{x}_C-\bm{x}_E}_2^2,\\
    &k_3=\kappa(8\pi+4)(1+\alpha^2)(y_E-y_C),\\
    & k_2=(1+\alpha^2)^2(x_E-x_C)^2+(1-\alpha^2)^2(y_E-y_C)^2\\
    &\qquad -2\kappa^2(2\pi+1)^2(1+\alpha^2),\\
    & k_1=\kappa(4\pi+2)(1-\alpha^2)^2(y_C-y_E),\\
    & k_0=\kappa^2(2\pi+1)^2(1-\alpha^2)^2.
    \end{aligned}
    \end{equation*}
\end{enumerate}
\end{thom}


\begin{proof}
The Lagrangian function for Problem \ref{prob:lowest-IP-lb} is
\begin{equation*}
\begin{aligned}
    &\mathcal{L}(\bm{x}_P^+,\bm{x}_E^+,\lambda_1,\lambda_2)=\alpha^2y_E^+-y_P^+-\alpha\enVert[1]{\bm{x}_P^+-\bm{x}_E^+}_2\\&\qquad
    +\lambda_1(\enVert[0]{\bm{x}_E^+-\bm{x}_E}_2-\frac{2\pi\kappa}{\alpha})+\lambda_2(\enVert[0]{\bm{x}_P^+-\bm{x}_C}_2-\kappa),
\end{aligned}
\end{equation*}
where $\lambda_i\in\mathbb{R}$ $(i=1,2)$ is the Lagrange multiplier. Let $(\bm{x}_P^*,\bm{x}^*_E)$ be an optimal solution to Problem \ref{prob:lowest-IP-lb}, leading to the optimal value $\widehat{\rho}_{\targetline}(X)=\alpha^2y_E^*-y_P^*-\alpha\enVert[0]{\bm{x}_P^*-\bm{x}_E^*}_2$. In view of the condition \ref{itm:nonconvex-pbm2-2} in Lemma \ref{lema:pbm2-relaxation}, $(\bm{x}_P^*,\bm{x}^*_E)$ lies at the boundary of the constraint set. The KKT condition for Problem \ref{prob:lowest-IP-lb} is
\begin{subequations}\label{eq:KKT-pbm2-1}
\begin{align}
0&=\pd{\mathcal{L}}{{x^+_E}}=\alpha\frac{x_P^\ast-x_E^\ast}{\enVert[1]{\bm{x}_P^\ast-\bm{x}_E^\ast}_2}+\lambda_1\frac{x_E^\ast-x_E}{\enVert[0]{\bm{x}_E^\ast-\bm{x}_E}_2},\label{subeq:KKT-pbm2-1}\\
0&=\pd{\mathcal{L}}{{y^+_E}}=\alpha\frac{y_P^\ast-y_E^\ast}{\enVert[1]{\bm{x}_P^\ast-\bm{x}_E^\ast}_2}+\lambda_1\frac{y_E^\ast-y_E}{\enVert[0]{\bm{x}_E^\ast-\bm{x}_E}_2}+\alpha^2,\label{subeq:KKT-pbm2-2}\\
0&=\pd{\mathcal{L}}{{x^+_P}}=\alpha\frac{x_E^\ast-x_P^\ast}{\enVert[1]{\bm{x}_P^\ast-\bm{x}_E^\ast}_2}+\lambda_2\frac{x_P^\ast-x_C}{\enVert[0]{\bm{x}_P^\ast-\bm{x}_C}_2},\label{subeq:KKT-pbm2-3}\\
0&=\pd{\mathcal{L}}{{y^+_P}}=\alpha\frac{y_E^\ast-y_P^\ast}{\enVert[1]{\bm{x}_P^\ast-\bm{x}_E^\ast}_2}+\lambda_2\frac{y_P^\ast-y_C}{\enVert[0]{\bm{x}_P^\ast-\bm{x}_C}_2}-1,\label{subeq:KKT-pbm2-4}
\end{align}
\end{subequations}
and 
\begin{equation}\label{eq:KKT-pbm2-2}
\enVert[1]{\bm{x}_E^\ast-\bm{x}_E}_2= \frac{2\pi\kappa}{\alpha},\ \enVert[1]{\bm{x}_P^\ast-\bm{x}_C}_2= \kappa,\ \lambda_1>0,\ \lambda_2>0.
\end{equation}
The reason for $\lambda_1\neq0$ and $\lambda_2\neq0$ is as follows. By \eqref{subeq:KKT-pbm2-2}, we have $\lambda_1\neq0$. If $\lambda_2=0$, then \eqref{subeq:KKT-pbm2-3} and \eqref{subeq:KKT-pbm2-4} lead to the contraction that $\textup{sgn}(y_E^*-y_P^*)\alpha-1=0$.

Regarding \ref{itm:solution-sign-pbm2}, consider the case $\sigma=\textup{sgn}(x_C-x_E)=1$, i.e., $x_C>x_E$ first. If $x_P^*=x_C$, then it follows from \eqref{subeq:KKT-pbm2-1} and \eqref{subeq:KKT-pbm2-3} that $x_C=x_E$. Thus, we have $x_P^*\neq x_C$. 

Assume that $x_P^*<x_C$. Let $\widetilde{\bm{x}}_P=[\widetilde{x}_P,\widetilde{y}_P]^\top=[2x_C-x_P^*,y_P^*]^\top$ and $\widetilde{\bm{x}}_E=[\widetilde{x}_E,\widetilde{y}_E]^\top=[x_E-\envert[0]{x_E-x_E^*},y_E^*]^\top$. Then, we have
\begin{equation*}
\enVert[0]{\widetilde{\bm{x}}_P-\bm{x}_C}_2\leq \kappa, \quad \enVert[0]{\widetilde{\bm{x}}_E-\bm{x}_E}_2\leq \frac{2\pi\kappa}{\alpha},
\end{equation*}
and 
\begin{equation*}
\begin{aligned}
\widetilde{x}_P-\widetilde{x}_E&=2x_C-x_P^*-x_E+\envert[0]{x_E-x_E^*}\\
&=\envert[0]{x_C-x_E}+\envert[0]{x_C-x_P^*}+\envert[0]{x_E-x_E^*}\\
&> \envert[0]{x_C-x_E-x_C+x_P^*+x_E-x_E^*}=\envert[0]{x_P^*-x_E^*}\\
\Rightarrow&\enVert[0]{\widetilde{\bm{x}}_P-\widetilde{\bm{x}}_E}_2>\enVert[0]{\bm{x}_P^*-\bm{x}_E^*}_2.
\end{aligned}
\end{equation*}
Thus, $(\widetilde{\bm{x}}_P,\widetilde{\bm{x}}_E)$ is a distinct feasible solution to Problem \ref{prob:lowest-IP-lb} and its induced value $\widetilde{\rho}$ satisfies
\begin{equation*}
\begin{aligned}
\widetilde{\rho}&=\alpha^2\widetilde{y}_E-\widetilde{y}_P-\alpha\enVert[0]{\widetilde{\bm{x}}_P-\widetilde{\bm{x}}_E}_2\\
&=\widehat{\rho}_{\mathcal{T}}(X)+\alpha\enVert[0]{\bm{x}_P^*-\bm{x}_E^*}_2-\alpha\enVert[0]{\widetilde{\bm{x}}_P-\widetilde{\bm{x}}_E}_2<\widehat{\rho}_{\mathcal{T}}(X),
\end{aligned}
\end{equation*}
which contradicts with the fact that $\widehat{\rho}_{\mathcal{T}}(X)$ is the optimal value. Therefore, $x_P^*> x_C$ holds, i.e., $\textup{sgn}(x_P^*-x_C)=\sigma$. As for the case $\sigma=\textup{sgn}(x_C-x_E)=-1$, the conclusion can be verified similarly. 

Next, consider the case $\sigma=\textup{sgn}(x_C-x_E)=0$, i.e., $x_C=x_E$. Suppose that $x_P^*\neq x_C$. Without loss of generality, let $x_P^*>x_C$. Then, according to \eqref{subeq:KKT-pbm2-1} and \eqref{subeq:KKT-pbm2-3}, we obtain that $x_P^*>x_C=x_E>x_E^*$. Let a unit vector $\bm{e}=[e_1,e_2]^\top$ such that $(\bm{x}_P^*-\bm{x}_E^*)^\top\bm{e}=0$. Thus, we have $e_2\neq 0$. Furthermore, we select $e_1$ such that starting from $\bm{x}_P$, $\bm{e}$ points into or is tangent with the constraint set of $\bm{x}^+_P$ such that it is feasible for $\bm{x}_P^+$ to move along the direction $\bm{e}$. By fixing $\bm{x}_E^+=\bm{x}_E^*$, the directional derivative of the objective function \eqref{subeq:pbm2-obj-func} along the vector $\bm{e}$ at $\bm{x}_P^*$ is
\begin{equation*}
\begin{aligned}
(\alpha^2-1)\nabla\rho_{\targetline}(\bm{x}_{I}^+)^\top\bm{e}=\alpha\frac{(\bm{x}_E^*-\bm{x}_P^*)^\top\bm{e}}{\enVert[1]{\bm{x}_P^\ast-\bm{x}_E^\ast}_2}-e_2=-e_2\neq 0,
\end{aligned}
\end{equation*}
which conflicts with the optimality of $(\bm{x}_P^*,\bm{x}_E^*)$. Therefore, we have $x_P^*=x_C$, i.e., $\textup{sgn}(x_P^*-x_C)=\sigma$. 

By following the same arguments, we also have $\textup{sgn}(x_E-x_E^*)=\sigma$. It follows from $\textup{sgn}(x^*_P-x_C)=\textup{sgn}(x_E-x_E^*)=\textup{sgn}(x_C-x_E)=\sigma$ that $\textup{sgn}(x_P^\ast-x_E^\ast)=\sigma$ holds.

Regarding \ref{itm:solution-pbm2}, we solve the solution $(\bm{x}_P^\ast,\bm{x}_E^\ast,\lambda_1,\lambda_2)$ to \eqref{eq:KKT-pbm2-1} and \eqref{eq:KKT-pbm2-2}. Combining \eqref{subeq:KKT-pbm2-1}, \eqref{subeq:KKT-pbm2-2} and \eqref{eq:KKT-pbm2-2} leads to
\begin{equation}\label{eq:yA-plus}
 \eqref{subeq:KKT-pbm2-1}^2 + \eqref{subeq:KKT-pbm2-2}^2 \Rightarrow y_E^\ast=y_E+\frac{(\alpha^2-\alpha^4-\lambda_1^2)\pi\kappa}{\alpha^3\lambda_1},
\end{equation}
and combining \eqref{subeq:KKT-pbm2-3}, \eqref{subeq:KKT-pbm2-4} and \eqref{eq:KKT-pbm2-2} leads to
\begin{equation}\label{eq:yD-plus}
\eqref{subeq:KKT-pbm2-3}^2 + \eqref{subeq:KKT-pbm2-4}^2 \Rightarrow y_P^\ast=y_C+\frac{(1-\alpha^2+\lambda_2^2)\kappa}{2\lambda_2}.
\end{equation}
Furthermore, it follows from \eqref{subeq:KKT-pbm2-2}, \eqref{subeq:KKT-pbm2-4} and \eqref{eq:KKT-pbm2-2} that
\begin{equation*}
\eqref{subeq:KKT-pbm2-2}+ \eqref{subeq:KKT-pbm2-4} \Rightarrow  \lambda_1\frac{\alpha(y_E^\ast-y_E)}{2\pi\kappa}+\lambda_2\frac{y_P^\ast-y_C}{\kappa}+\alpha^2-1=0.
\end{equation*}
Thus, substituting \eqref{eq:yA-plus} and \eqref{eq:yD-plus} into the above equation, we have $\lambda_1=\alpha\lambda_2$. 
According to the sign relationship \ref{itm:solution-sign-pbm2}, then \eqref{eq:KKT-pbm2-2} and \eqref{eq:yD-plus} lead to
\begin{equation}\label{eq:xD-plus}
    x_P^\ast=x_C+\frac{\sigma\kappa}{2\lambda_2}\phi(\lambda_2),
\end{equation}
where $\phi(\lambda_2)=\sqrt{2(1+\alpha^2)\lambda_2^2-\lambda_2^4-(1-\alpha^2)^2}$. Furthermore, it follows from \eqref{subeq:KKT-pbm2-1}, \eqref{subeq:KKT-pbm2-3}, \eqref{eq:KKT-pbm2-2} and \eqref{eq:xD-plus} that
\begin{equation}\label{eq:xA-plus}
\begin{aligned}
\eqref{subeq:KKT-pbm2-1} +  \eqref{subeq:KKT-pbm2-3} \Rightarrow x_E^\ast&=x_E-\frac{2\pi\lambda_2(x_P^\ast-x_C)}{\alpha\lambda_1}\\
&=x_E-\frac{\sigma\pi\kappa}{\alpha^2\lambda_2}\phi.
\end{aligned}
\end{equation}

Then, by \eqref{subeq:KKT-pbm2-3}, \eqref{eq:KKT-pbm2-2} and \eqref{eq:xD-plus}, we obtain
\begin{equation}\label{eq:xA-xD-yA-yD-1}
\begin{aligned}
   &\frac{\alpha^2(x_E^\ast-x_P^\ast)^2}{\enVert[0]{\bm{x}_P^\ast-\bm{x}_E^\ast}_2^2}=\frac{\lambda_2^2(x_P^\ast-x_C)^2}{\kappa^2}=\frac{\phi^2}{4}\\
  & \Rightarrow (x_E^\ast-x_P^\ast)^2(4\alpha^2-\phi^2)=(y_E^\ast-y_P^\ast)^2 \phi^2.
\end{aligned}
\end{equation}
Note that 
\begin{equation*}
\begin{aligned}
4\alpha^2-\phi^2&=4\alpha^2-2(1+\alpha^2)\lambda_2^2+\lambda_2^4+(1-\alpha^2)^2\\
&=(1+\alpha^2-\lambda_2^2)^2.
\end{aligned}
\end{equation*}
Thus, \eqref{eq:xA-xD-yA-yD-1} becomes 
\begin{equation}\label{eq:abs-phi-lambda}
   \envert[1]{(x_P^\ast-x_E^\ast)(1+\alpha^2-\lambda_2^2)}= \envert[1]{y_E^\ast-y_P^\ast}\phi.
\end{equation}
Note that by \eqref{eq:yD-plus}, \eqref{subeq:KKT-pbm2-4} implies that
\begin{equation*}
\begin{aligned}
&\alpha\frac{y_E^\ast-y_P^\ast}{\enVert[1]{\bm{x}_P^\ast-\bm{x}_E^\ast}_2}=1-\lambda_2\frac{y_P^\ast-y_C}{\enVert[0]{\bm{x}_P^\ast-\bm{x}_C}_2}\\
&\qquad\qquad \qquad\hspace{0.09cm} =1-\lambda_2\frac{y_P^\ast-y_C}{\kappa}=\frac{1+\alpha^2-\lambda_2^2}{2}\\
&\Rightarrow \textup{sgn}(y_E^\ast-y_P^\ast)=\textup{sgn}(1+\alpha^2-\lambda_2^2).
\end{aligned}
\end{equation*}
From the condition \ref{itm:solution-sign-pbm2}, we have $\textup{sgn}(x_P^\ast-x_E^\ast)=\sigma$. Thus, \eqref{eq:abs-phi-lambda} can be simplified as follows
\begin{equation*}
\begin{aligned}
 &\sigma(x_P^\ast-x_E^\ast)(1+\alpha^2-\lambda_2^2)= (y_E^\ast-y_P^\ast)\phi\\
 &\Rightarrow\big((y_E-y_C)\lambda_2-(2\pi+1)\kappa\big)\phi \\
 &\qquad =\sigma(x_E-x_C)\big(\lambda_2^3-(1+\alpha^2)\lambda_2\big),
\end{aligned}
\end{equation*}
where \eqref{eq:yA-plus}, \eqref{eq:yD-plus}, \eqref{eq:xD-plus}, \eqref{eq:xA-plus} and $\lambda_1=\alpha\lambda_2$ are used. Substituting the expression of $\phi$ into the above equation, we obtain that $\lambda_2$ is a positive solution of the sextic equation \eqref{eq:lambda-six}. Thus, according to \eqref{eq:yA-plus}, \eqref{eq:yD-plus}, \eqref{eq:xD-plus} and \eqref{eq:xA-plus}, the conclusion can be obtained.
\end{proof}

Now we present the main result in this section, providing a pursuit winning strategy for SC and Non-IO, for which a sufficient and efficient capture condition is given.

\begin{thom}[Pursuit winning strategy for SC and Non-IO]\label{thom:dwin-stra-no-IO}
Consider $P\in\dteam$ and $E\in\ateam$ under the model \eqref{eq:pursuer_car} and \eqref{eq:evader_simple}, respectively. Let a state $X=(\bm{x}_{P},\theta_{P},\bm{x}_{E})$ such that
\begin{enumerate}[label=(\roman*)]
    \item $X$ satisfies SC;
    \item the conditions \ref{itm:no-capture}-\ref{item:no-close} in Lemma \ref{lema:Heading-modi-time} are true.
\end{enumerate}
Then, if the parameters $(r,\kappa,\alpha)$ satisfy
\begin{equation}\label{eq:t-k-max}
\begin{aligned}
  \frac{r}{\kappa}>\max\Big\{h(\alpha),\frac{(\alpha+1)^2}{\alpha(\alpha-1)}\Big\},
\end{aligned}
\end{equation}
and the optimal value $\widehat{\rho}_{\mathcal{T}}(X)\ge0$, then $P$ can adopt the feedback strategy \eqref{eq:heading-strategy} such that the IO holds after a finite time, and afterwards $P$ can adopt the feedback strategy \eqref{eq:dwincontrolzero} to guarantee $\rho(\CAR,\goal)\ge0$ and \eqref{eq:odzeroradius}, regardless of $\bm{u}_{E}\in\mathbb{S}^1$.
\end{thom}
\begin{proof}
Note that $X$ satisfies SC but does not satisfies IO. We propose a two-step strategy for the pursuer and give a sufficient condition to guarantee the winning under this strategy. The first step is to adjust its heading by the heading adjustment strategy \eqref{eq:heading-strategy} such that the IO holds after a finite time, as Fig. \ref{fig:heading-adjustment} shows. Then, the second step is to adopt the strategy \eqref{eq:dwincontrolzero} which has been proved to have the winning guarantee under some conditions when the SC and IO both hold. The key point in this two-step strategy is to determine whether the pursuer and the evader satisfy the pursuer's winning conditions for the second step once the pursuer completes the first step. This question can be resolved by the solution to \pbmref{prob:lowest-IP}, which is quite complex.  We are able to find a lower bound of the optimal value of  \pbmref{prob:lowest-IP} by solving a simpler \pbmref{prob:lowest-IP-lb} for which we have proposed an efficient solution in \thomref{thom:solution-pbm-2}. 

Under this two-step strategy, the winning capture condition is straightforward by combining Theorems \ref{thom:dwinstra-car-zero}, \ref{thom:heading-adjustment} and the conclusion \ref{itm:nonconvex-pbm2-3} in Lemma \ref{lema:pbm2-relaxation}. 
\end{proof}

\begin{rek}
Since \lemaref{lema:sufficient-condition-IO} has provided an upper bound of $h(\alpha)$, we can first check the following condition before \eqref{eq:t-k-max}:
\begin{equation*}
\begin{aligned}
  \frac{r}{\kappa}>\max\Big\{\frac{2\alpha-1}{(\alpha-1)^2},\frac{(\alpha+1)^2}{\alpha(\alpha-1)}\Big\},
\end{aligned}
\end{equation*}
which is easy to verify. Furthermore, if $\alpha\ge\alpha_0$, the second part is the maximum; otherwise, the first part is the maximum, where $\alpha_0\in\mathbb{R}_{>0}$ is the unique positive solution of $\alpha^3-\alpha^2-1=0$. If $\alpha<\alpha_0$ and the above condition fails, then we have to turn to the condition \eqref{eq:t-k-max}.
\end{rek}



\section{Multiplayer Games}\label{sec:multiplayer}
We piece together the pairwise outcomes of all pursuers and evaders using maximum matching for the task assignment of pursuers as in \cite{RY-XD-ZS-YZ-FB:19,MC-ZZ-CJT:17}.  Then, a receding horizon pursuit strategy is proposed, allowing to capture more evaders as the game runs. This strategy is indeed useful because the cooperation among pursuers in the strategy level is not considered and thus a better matching may occur as the game evolves.

Let $\graph=(\dteam\cup \ateam,\edgeset)$ be an undirected bipartite graph consisting of two node sets $\dteam$ and $\ateam$, where $\edgeset$ is a set of edges. Denote by $e_{ij}$ the edge connecting node $P_i\in \dteam$ and node $E_j\in\ateam$. An edge $e_{ij}\in\edgeset$ if and only if $P_i$ can win against $E_j$ by Theorems \ref{thom:dwinstra-car-zero} or \ref{thom:dwin-stra-no-IO}, depending on whether the IO is satisfied.

The receding horizon pursuit strategy is as follows. At time $t$, the current states of all one-pursuer-one-evader subgames are used to construct the graph $\graph$. Then, a maximum matching $M$ of $\graph$ is computed. Each pursuer in $M$ is assigned to the matched evader, and adopts the guaranteed winning strategy \eqref{eq:dwincontrolzero} or the combination of \eqref{eq:heading-strategy} and \eqref{eq:dwincontrolzero}, depending on whether the IO holds. The pursuers which are not matched in $M$ will pursue one of unmatched evaders. Each player updates its position (and heading) after a small constant time step. The new states are then used to construct a new graph, and the process iterates until no evaders in $\play$.

\section{Simulation Results}\label{sec:simulation}

\begin{figure}[tb]
    \centering
    \includegraphics[width=0.95\hsize]{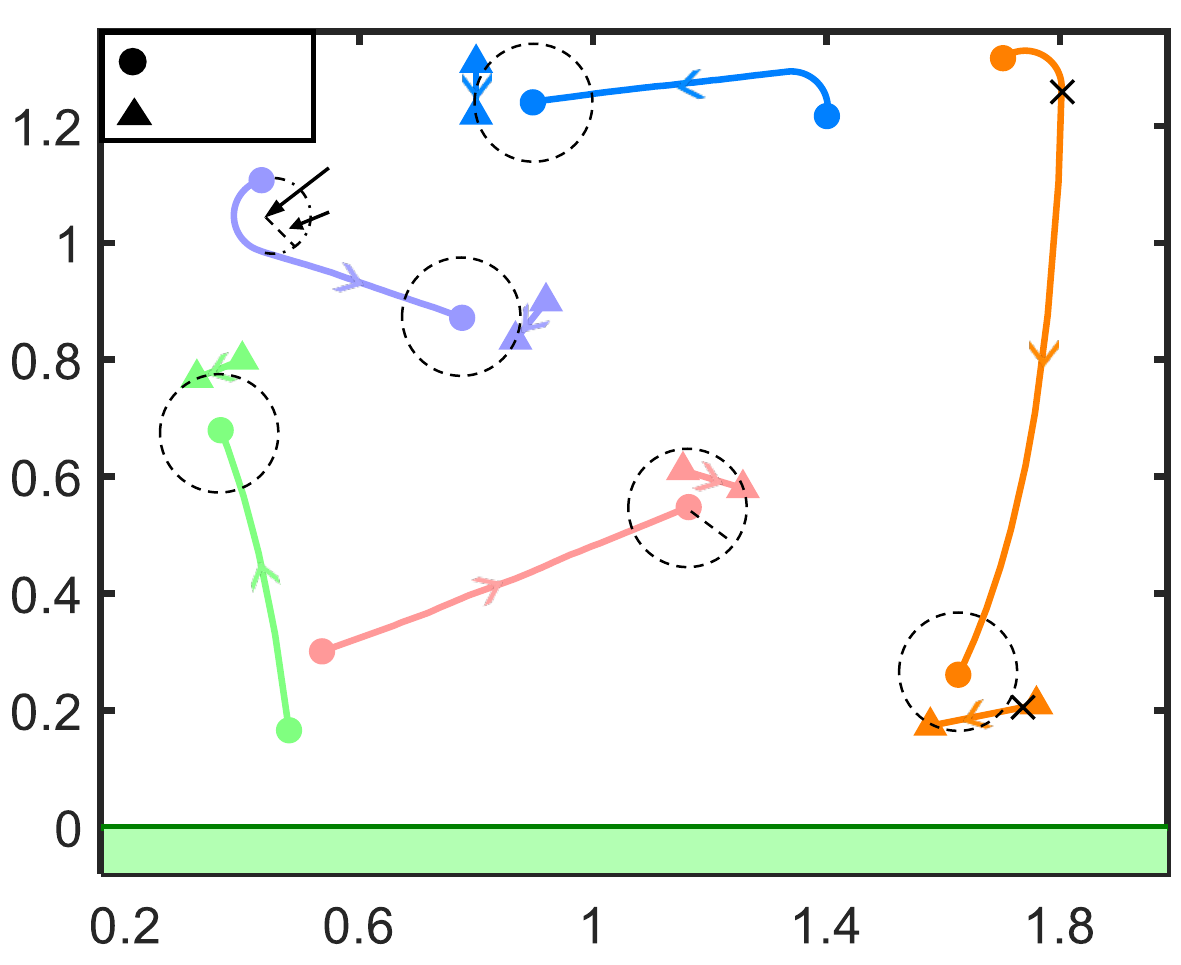}
    \put(-205,181){\footnotesize pursuers}
    \put(-205,171){\footnotesize evaders}
    \put(-206,159){\footnotesize $\bm{x}_{P_1}$}
    \put(-44,175){\footnotesize $\bm{x}_{P_2}$}
    \put(-199,45){\footnotesize $\bm{x}_{P_3}$}
    \put(-178,55){\footnotesize $\bm{x}_{P_4}$}
    \put(-78,163){\footnotesize $\bm{x}_{P_5}$}
    \put(-162,182){\footnotesize $\bm{x}_{E_1}$}
    \put(-106,109){\footnotesize $\bm{x}_{E_2}$}
    \put(-192,129){\footnotesize $\bm{x}_{E_3}$}
    \put(-130,142){\footnotesize $\bm{x}_{E_4}$}
    \put(-25,55){\footnotesize $\bm{x}_{E_5}$}
    \put(-101,84){\footnotesize $r$}
    \put(-174,163){\footnotesize $\bm{x}_C$}
    \put(-171,151){\footnotesize $\kappa$}
    \put(-10,8){\footnotesize $x$}
    \put(-230,185){\footnotesize $y$}
    \put(-26,34){\footnotesize $\play$}
    \put(-26,21){\footnotesize $\goal$}
    \caption{Simulation for a multiplayer game with five pursuers $P_i$ ($i=1,\dots,5$, solid circles) and five evaders $E_j$ ($j=1,\dots,5$, solid triangles). The trajectories of the matched pursuer and evader are in the same color, with arrows indicating their moving directions. The dashed circles are the boundaries of the capture regions when the evaders are captured. The dash dot line is the minimum turning radius for $P_1$.}
    \label{fig:sim_5v5}
\end{figure}
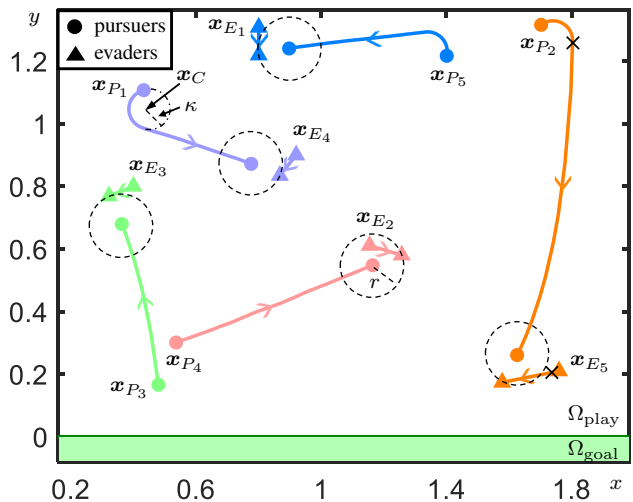

To illustrate the proposed guaranteed winning strategies for pursuers in multiplayer reach-avoid games,  a simulation of the game with five pursuers and five evaders is shown in Fig. \ref{fig:sim_5v5}. In the game, pursuers are modeled as Dubins cars \eqref{eq:pursuer_car} and evaders have simple-motion \eqref{eq:evader_simple}. We assume that all pursuers have the same minimum turning radius, capture radius and speed, and all evaders have the same speed. The strategies still apply to heterogeneous players. We consider the parameters $r=0.1, \kappa=0.0625,v_P=0.3$ and $\alpha=6.3$ which satisfy the parameter relations \eqref{eq:parameterrelazero} and \eqref{eq:t-k-max} for the case of SC and IO and the case of SC and Non-IO, respectively. The trajectories of the matched pursuer and evader are in the same color, with arrows indicating their moving directions. The dashed circles are the boundaries of the capture regions. The dash dot line is the minimum turning radius for $P_1$.

Three different cases are included in this example: $(a)$ SC and IO; $(b)$ SC and Non-IO; $(c)$ no guaranteed pursuit winning initially but later have one. The pursuer-evader pairs $(P_3,E_3)$ and $(P_4,E_2)$ satisfy SC and IO in Theorem \ref{thom:dwinstra-car-zero} initially, and each pursuer adopts the feedback strategy \eqref{eq:dwincontrolzero}. The pairs $(P_1,E_4)$ and $(P_5,E_1)$ satisfy SC and Non-IO, as well as the conditions in Theorem \ref{thom:dwin-stra-no-IO} initially. Each pursuer adopts the two-step strategy in Theorem \ref{thom:dwin-stra-no-IO}. Therefore, $P_3,P_4,P_1$ and $P_5$ can guarantee to win against the matched evaders. For the pair $(P_2,E_5)$, the conditions in Theorems \ref{thom:dwinstra-car-zero} and \ref{thom:dwin-stra-no-IO} do not hold, so we cannot tell a guaranteed pursuit winning at this time. Thus, after the task assignment, $P_2$ and $E_5$ are unmatched. However, as $P_2$ adopts strategy \eqref{eq:heading-strategy} and the game evolves, the SC and IO with respect to $E_5$ are achieved at the place marked by black crosses. Afterwards, $P_2$ turns to the strategy \eqref{eq:dwincontrolzero} and finally captures $E_5$. The strategies of all evaders are selected randomly conditioned that each evader moves along a straight line and gets closer to the goal region $\goal$.

\section{Conclusion}\label{sec:conclusion}
We presented an analytical approach to multiplayer Homicidal Chauffeur reach-avoid differential games in which pursuers protect a goal region from evaders. We considered Dubins-car pursuers and simple-motion evaders, which is a more accurate model and has broad applications. For the subgame with simple-motion pursuer, the proposed pursuit feedback strategy can guarantee the ER not close to the goal region if they are separate initially (i.e., the SC holds), thus leading to a pursuit winning. Based on it, if the SC holds, we proposed two feedback strategies for the Dubins-car pursuer in the IO and Non-IO cases respectively, and presented the respective conditions on player configurations and parameters to guarantee the pursuer's winning. To ease the computation burden, an efficient sufficient condition on parameters for the pursuer's winning in the IO case was given, and a relaxation problem which can be solved efficiently, was proposed to determine the  pursuer's winning sufficiently in the Non-IO case. A strategy for the pursuit team was proposed by piecing together the subgame outcomes with maximum matching and adopting the receding horizon method. Future work will involve reversed Homicidal Chauffeur reach-avoid games and distributed reach-avoid games.

\bibliographystyle{plainurl} 
\bibliography{reference}

\begin{thebibliography}{10}

\bibitem{TB-GJO:99}
T.~Basar and G.~J. Olsder.
\newblock {\em Dynamic Noncooperative Game Theory}.
\newblock SIAM, 1999.

\bibitem{SPB-DSB:00}
S.~P. Bhat and D.~S. Bernstein.
\newblock Finite-time stability of continuous autonomous systems.
\newblock {\em SIAM Journal on Control and Optimization}, 38(3):751--766, 2000.

\bibitem{SDB-FB-JPH:09}
S.~D. Bopardikar, F.~Bullo, and J.~P. Hespanha.
\newblock A cooperative {H}omicidal {C}hauffeur game.
\newblock {\em Automatica}, 45(7):1771--1777, 2009.
\newblock \href
  {https://doi.org/https://doi.org/10.1016/j.automatica.2009.03.014}
  {\path{doi:https://doi.org/10.1016/j.automatica.2009.03.014}}.

\bibitem{MC-SB-JFF-CJT:19}
M.~{Chen}, S.~{Bansal}, J.~F. {Fisac}, and C.~J. {Tomlin}.
\newblock Robust sequential trajectory planning under disturbances and
  adversarial intruder.
\newblock {\em IEEE Transactions on Control Systems Technology},
  27(4):1566--1582, 2019.
\newblock \href {https://doi.org/10.1109/TCST.2018.2828380}
  {\path{doi:10.1109/TCST.2018.2828380}}.

\bibitem{MC-SLH-MSV-SB-CJT:18}
M.~{Chen}, S.~L. {Herbert}, M.~S. {Vashishtha}, S.~{Bansal}, and C.~J.
  {Tomlin}.
\newblock Decomposition of reachable sets and tubes for a class of nonlinear
  systems.
\newblock {\em IEEE Transactions on Automatic Control}, 63(11):3675--3688,
  2018.
\newblock \href {https://doi.org/10.1109/TAC.2018.2797194}
  {\path{doi:10.1109/TAC.2018.2797194}}.

\bibitem{MC-ZZ-CJT:17}
M.~{Chen}, Z.~{Zhou}, and C.~J. {Tomlin}.
\newblock Multiplayer reach-avoid games via pairwise outcomes.
\newblock {\em IEEE Transactions on Automatic Control}, 62(3):1451--1457, 2017.
\newblock \href {https://doi.org/10.1109/TAC.2016.2577619}
  {\path{doi:10.1109/TAC.2016.2577619}}.

\bibitem{SC-MP-RM:17}
S.~Coates, M.~Pachter, and R.~Murphey.
\newblock Optimal control of a {D}ubins car with a capture set and the
  {H}omicidal {C}hauffeur differential game.
\newblock {\em IFAC-PapersOnLine}, 50(1):5091--5096, 2017.
\newblock \href {https://doi.org/10.1016/j.ifacol.2017.08.775}
  {\path{doi:10.1016/j.ifacol.2017.08.775}}.

\bibitem{RJE-NJK:72}
R.~J. Elliott and N.~J. Kalton.
\newblock {\em The existence of value in differential games}.
\newblock American Mathematical Soc., 1972.

\bibitem{IE-PT-MP:15}
I.~Exarchos, P.~Tsiotras, and M.~Pachter.
\newblock On the suicidal pedestrian differential game.
\newblock {\em Dynamic Games and Applications}, 5(3):297--317, 2015.
\newblock \href {https://doi.org/10.1007/s13235-014-0130-2}
  {\path{doi:10.1007/s13235-014-0130-2}}.

\bibitem{JFF-MC-CJT-SSS:15}
J.~F. Fisac, M.~Chen, C.~J. Tomlin, and S.~S. Sastry.
\newblock Reach-avoid problems with time-varying dynamics, targets and
  constraints.
\newblock In {\em Proceedings of the 18th International Conference on Hybrid
  Systems: Computation and Control}, page 11–20, Seattle, Washington, 2015.
\newblock \href {https://doi.org/10.1145/2728606.2728612}
  {\path{doi:10.1145/2728606.2728612}}.

\bibitem{EG-DWC-MP:20}
E.~{Garcia}, D.~W. {Casbeer}, and M.~{Pachter}.
\newblock Optimal strategies for a class of multi-player reach-avoid
  differential games in 3{D} space.
\newblock {\em IEEE Robotics and Automation Letters}, 5(3):4257--4264, 2020.
\newblock \href {https://doi.org/10.1109/LRA.2020.2994023}
  {\path{doi:10.1109/LRA.2020.2994023}}.

\bibitem{EG-DWC-MP:20-2}
E.~{Garcia}, D.~W. {Casbeer}, and M.~{Pachter}.
\newblock Optimal strategies of the differential game in a circular region.
\newblock {\em IEEE Control Systems Letters}, 4(2):492--497, 2020.
\newblock \href {https://doi.org/10.1109/LCSYS.2019.2963173}
  {\path{doi:10.1109/LCSYS.2019.2963173}}.

\bibitem{EG-DWC-AVM-MP:20}
E.~{Garcia}, D.~W. {Casbeer}, A.~{Von Moll}, and M.~{Pachter}.
\newblock Multiple pursuer multiple evader differential games.
\newblock {\em IEEE Transactions on Automatic Control}, pages 1--1, 2020.
\newblock \href {https://doi.org/10.1109/TAC.2020.3003840}
  {\path{doi:10.1109/TAC.2020.3003840}}.

\bibitem{AG-EA-SS:17}
A.~Gray, E.~Abbena, and S.~Salamon.
\newblock {\em Modern differential geometry of curves and surfaces with
  Mathematica}.
\newblock Chapman and Hall/CRC, 2017.

\bibitem{RI:65}
R.~Isaacs.
\newblock {\em Differential Games}.
\newblock New York: Wiley, 1965.

\bibitem{JL:12}
J.~Lewin.
\newblock {\em Differential games: theory and methods for solving game problems
  with singular surfaces}.
\newblock Springer Science \& Business Media, 2012.

\bibitem{HL-YW-FLW-KPV:20}
H.~{Liu}, Y.~{Wang}, F.~L. {Lewis}, and K.~P. {Valavanis}.
\newblock Robust formation tracking control for multiple quadrotors subject to
  switching topologies.
\newblock {\em IEEE Transactions on Control of Network Systems},
  7(3):1319--1329, 2020.
\newblock \href {https://doi.org/10.1109/TCNS.2020.2976271}
  {\path{doi:10.1109/TCNS.2020.2976271}}.

\bibitem{KM-JL:11}
K.~{Margellos} and J.~{Lygeros}.
\newblock Hamilton–{J}acobi formulation for reach–avoid differential games.
\newblock {\em IEEE Transactions on Automatic Control}, 56(8):1849--1861, 2011.
\newblock \href {https://doi.org/10.1109/TAC.2011.2105730}
  {\path{doi:10.1109/TAC.2011.2105730}}.

\bibitem{AWM:71}
A.~W. Merz.
\newblock {\em The homicidal chauffeur--a differential game}.
\newblock PhD thesis, 1971.

\bibitem{IMM-AMB-CJT:05}
I.~M. {Mitchell}, A.~M. {Bayen}, and C.~J. {Tomlin}.
\newblock A time-dependent {H}amilton-{J}acobi formulation of reachable sets
  for continuous dynamic games.
\newblock {\em IEEE Transactions on Automatic Control}, 50(7):947--957, 2005.
\newblock \href {https://doi.org/10.1109/TAC.2005.851439}
  {\path{doi:10.1109/TAC.2005.851439}}.

\bibitem{DWO-ARG:16}
D.~W. {Oyler} and A.~R. {Girard}.
\newblock Dominance regions in the {H}omicidal {C}hauffeur problem.
\newblock In {\em 2016 American Control Conference (ACC)}, pages 2494--2499,
  2016.
\newblock \href {https://doi.org/10.1109/ACC.2016.7525291}
  {\path{doi:10.1109/ACC.2016.7525291}}.

\bibitem{MP-SC:19}
M.~Pachter and S.~Coates.
\newblock The classical homicidal chauffeur game.
\newblock {\em Dynamic Games and Applications}, 9(3):800--850, 2019.
\newblock \href {https://doi.org/10.1007/s13235-018-0264-8}
  {\path{doi:10.1007/s13235-018-0264-8}}.

\bibitem{VSP-VLT:11}
V.~S. Patsko and V.~L. Turova.
\newblock {\em Homicidal Chauffeur Game: History and Modern Studies}, pages
  227--251.
\newblock 2011.
\newblock \href {https://doi.org/10.1007/978-0-8176-8089-3_12}
  {\path{doi:10.1007/978-0-8176-8089-3_12}}.

\bibitem{UR-RM:16}
U.~Ruiz and R.~Murrieta-Cid.
\newblock A differential pursuit/evasion game of capture between an
  omnidirectional agent and a differential drive robot, and their winning
  roles.
\newblock {\em International Journal of Control}, 89(11):2169--2184, 2016.
\newblock \href {https://doi.org/10.1080/00207179.2016.1151078}
  {\path{doi:10.1080/00207179.2016.1151078}}.

\bibitem{DS-VK:20}
D.~Shishika and V.~Kumar.
\newblock A review of multi agent perimeter defense games.
\newblock In {\em International Conference on Decision and Game Theory for
  Security}, pages 472--485, 2020.
\newblock \href {https://doi.org/10.1007/978-3-030-64793-3_26}
  {\path{doi:10.1007/978-3-030-64793-3_26}}.

\bibitem{VMA-EG-DC-MS-SCS:20}
A.~Von~Moll, E.~Garcia, D.~Casbeer, M.~Suresh, and S.~C. Swar.
\newblock Multiple-pursuer, single-evader border defense differential game.
\newblock {\em Journal of Aerospace Information Systems}, 17(8):407--416, 2020.
\newblock \href {https://doi.org/10.2514/1.I010740}
  {\path{doi:10.2514/1.I010740}}.

\bibitem{RY-XD-ZS-YZ-FB:19}
R.~Yan, X.~Duan, Z.~Shi, Y.~Zhong, and F.~Bullo.
\newblock Matching-based capture strategies for 3{D} heterogeneous multiplayer
  reach-avoid differential games.
\newblock {\em arXiv preprint:1909.11881}, 2019.

\bibitem{RY-ZS-YZ:19-2}
R.~Yan, Z.~Shi, and Y.~Zhong.
\newblock Optimal strategies for the lifeline differential game with limited
  lifetime.
\newblock {\em International Journal of Control}, pages 1--14, 2019.
\newblock \href {https://doi.org/10.1080/00207179.2019.1698770}
  {\path{doi:10.1080/00207179.2019.1698770}}.

\bibitem{RY-ZS-YZ:19}
R.~{Yan}, Z.~{Shi}, and Y.~{Zhong}.
\newblock Reach-avoid games with two defenders and one attacker: An analytical
  approach.
\newblock {\em IEEE Transactions on Cybernetics}, 49(3):1035--1046, 2019.
\newblock \href {https://doi.org/10.1109/TCYB.2018.2794769}
  {\path{doi:10.1109/TCYB.2018.2794769}}.

\bibitem{RY-ZS-YZ:20-2}
R.~{Yan}, Z.~{Shi}, and Y.~{Zhong}.
\newblock Guarding a subspace in high-dimensional space with two defenders and
  one attacker.
\newblock {\em IEEE Transactions on Cybernetics}, pages 1--14, 2020.
\newblock \href {https://doi.org/10.1109/TCYB.2020.3015031}
  {\path{doi:10.1109/TCYB.2020.3015031}}.

\bibitem{RY-ZS-YZ:20-1}
R.~{Yan}, Z.~{Shi}, and Y.~{Zhong}.
\newblock Task assignment for multiplayer reach–avoid games in convex domains
  via analytical barriers.
\newblock {\em IEEE Transactions on Robotics}, 36(1):107--124, 2020.
\newblock \href {https://doi.org/10.1109/TRO.2019.2935345}
  {\path{doi:10.1109/TRO.2019.2935345}}.

\bibitem{JY-XD-QL-JL-ZR:19}
J.~{Yu}, X.~{Dong}, Q.~{Li}, J.~{Lü}, and Z.~{Ren}.
\newblock Fully adaptive practical time-varying output formation tracking for
  high-order nonlinear stochastic multiagent system with multiple leaders.
\newblock {\em IEEE Transactions on Cybernetics}, pages 1--13, 2019.
\newblock \href {https://doi.org/10.1109/TCYB.2019.2956316}
  {\path{doi:10.1109/TCYB.2019.2956316}}.

\bibitem{JY-XD-QL-ZR:18}
J.~{Yu}, X.~{Dong}, Q.~{Li}, and Z.~{Ren}.
\newblock Practical time-varying formation tracking for second-order nonlinear
  multiagent systems with multiple leaders using adaptive neural networks.
\newblock {\em IEEE Transactions on Neural Networks and Learning Systems},
  29(12):6015--6025, 2018.
\newblock \href {https://doi.org/10.1109/TNNLS.2018.2817880}
  {\path{doi:10.1109/TNNLS.2018.2817880}}.

\bibitem{ZZ-RT-HH-CJT:12}
Z.~{Zhou}, R.~{Takei}, H.~{Huang}, and C.~J. {Tomlin}.
\newblock A general, open-loop formulation for reach-avoid games.
\newblock In {\em 2012 IEEE 51st IEEE Conference on Decision and Control
  (CDC)}, pages 6501--6506, 2012.
\newblock \href {https://doi.org/10.1109/CDC.2012.6426643}
  {\path{doi:10.1109/CDC.2012.6426643}}.

\end{thebibliography}

\end{document}